\documentclass[journal]{IEEEtran}
\usepackage{cite}
\usepackage[pdftex]{graphicx}
\usepackage[cmex10]{mathtools}
\usepackage{amssymb}
\usepackage[usenames,dvipsnames]{color}
\usepackage{array}
\usepackage[caption=false,font=footnotesize]{subfig}

\usepackage{bm}
\usepackage{bbm}
\usepackage{microtype}

 \def\cB{{\mathcal{B}}} \def\cC{{\mathcal{C}}} 
\def\cE{{\mathcal{E}}}

\def\ba{{\mathbf{a}}}  \def\bc{{\mathbf{c}}} 
\def\bee{{\mathbf{e}}} \def\bff{{\mathbf{f}}}  \def\bh{{\mathbf{h}}}
   
 \def\bn{{\mathbf{n}}}  
   
 \def\bv{{\mathbf{v}}} \def\bw{{\mathbf{w}}} 
\def\by{{\mathbf{y}}} \def\bz{{\mathbf{z}}}  

\def\bA{{\mathbf{A}}}  \def\bC{{\mathbf{C}}} 
  \def\bG{{\mathbf{G}}} \def\bH{{\mathbf{H}}}
\def\bI{{\mathbf{I}}}   
 \def\bN{{\mathbf{N}}}  
   
\def\bU{{\mathbf{U}}} \def\bV{{\mathbf{V}}}  \def\bX{{\mathbf{X}}}
 \def\bZ{{\mathbf{Z}}}

\DeclareMathOperator*{\argmin}{arg\,min}
\DeclareMathOperator*{\argmax}{arg\,max}

\begin{document}


\title{Adaptive Millimeter Wave Beam Alignment for Dual-Polarized MIMO Systems}


\author{Jiho Song, Junil Choi, Stephen G. Larew, David J. Love, Timothy A. Thomas, and Amitava Ghosh
\thanks{J.\ Song, J.\ Choi, S.\ G.\ Larew, and D.\ J.\ Love are with the School of Electrical and Computer Engineering, Purdue University, West Lafayette, IN 47907 (e-mail: \{jihosong, choi215, sglarew, djlove\}@purdue.edu).}
\thanks{T.\ A.\ Thomas and A.\ Ghosh are with Nokia Solutions and Networks, Arlington Heights, IL 60004 (e-mail: \{timothy.thomas, amitava.ghosh\}@nsn.com).}
}




\maketitle


\begin{abstract}
Fifth generation wireless systems are expected to employ multiple antenna communication at millimeter wave (mmWave) frequencies using small cells within heterogeneous cellular networks. The high path loss of mmWave as well as physical obstructions make communication challenging. To compensate for the severe path loss, mmWave systems may employ a beam alignment algorithm that facilitates highly directional transmission by aligning the beam direction of multiple antenna arrays. This paper discusses a mmWave system employing dual-polarized antennas. First, we propose a practical soft-decision beam alignment (soft-alignment) algorithm that exploits orthogonal polarizations. By sounding the orthogonal polarizations in parallel, the equality criterion of the Welch bound for training sequences is relaxed.  Second, the analog beamforming system is adapted to {the directional characteristics} of the mmWave link assuming a high Ricean ${K}$-factor and poor scattering environment. The soft-algorithm enables the mmWave system to align {innumerable narrow beams to channel subspace} in an attempt to effectively scan the mmWave channel. Thirds, we propose a method to efficiently adapt the number of channel sounding observations to the specific channel environment based on an approximate probability of beam misalignment. Simulation results show the proposed soft-alignment algorithm with adaptive sounding time effectively scans the channel subspace of a mobile user by exploiting polarization diversity.
\end{abstract}


\IEEEpeerreviewmaketitle



\section{Introduction}
\label{sec:Introduction}
\IEEEPARstart{M}{illimeter} wave (mmWave) multiple-input multiple-output (MIMO) systems are a prime candidate to allow future communication systems to provide the throughput enhancements needed to meet the expected demands for mobile data~\cite{Ref_Gho14, Ref_Pi11,Ref_Rap13,Ref_Roh14}.  Radio links operating over the wide bandwidths available in the mmWave spectrum are a promising method of providing access for small cells within heterogeneous cellular networks. When compared to lower frequencies, the higher expected path loss of mmWave requires greater system gains in the link budget \cite{Ref_Pi11}, which may be attained via beamforming with multiple antenna systems. The small wavelength of mmWaves allows for a dense packing of many antennas in a small space. In effect, a steerable, phased-array architecture of many antennas can be controlled to form high-gain directional transmissions. 
However, mmWave systems may have only a small number of radio frequency (RF) chains due to their high cost and power utilization. Therefore, beamforming in mmWave systems is performed mainly by using inexpensive RF phase shifters in the analog domain~\cite{Ref_Roh14}.

In contrast to conventional cellular systems using one RF chain per antenna, analog beamformed or precoded mmWave systems cannot observe the channel of each receive antenna directly since the incident waves at each antenna are combined in the analog domain. Moreover, the large number of antennas and the high path loss at mmWave frequencies make it difficult to acquire enough samples to compute meaningful channel estimates for each receive antenna. Instead, one approach is to perform subspace sampling using a finite number of training vectors for beam alignment. The training vectors are designed to sound the mmWave channel in an attempt to align the beamformers to the channel subspace.

Several beam alignment techniques for mmWave systems have been developed based on hard-decision beam alignment (hard-alignment)  techniques \cite{Ref_Hur13,Ref_Hur11,Ref_IEEE802.11,Ref_Son13,Ref_Son13_2}. In the hard-alignment algorithms, a set of candidate training vectors are used to scan the channel subspace. The chosen beamformer for data transmission comes from the same set used for channel subspace scanning. Hard-alignment sampling algorithms utilizing a hierarchical multi-round beam search approach are studied for uniform linear array (ULA) scenarios in \cite{Ref_Hur13, Ref_Hur11} and extended to dual-polarized arrays in \cite{Ref_Son13, Ref_Son13_2}, which show better performance than the standardized technique in \cite{Ref_IEEE802.11}. However, the hard-alignment techniques have fundamental limits on beam alignment performance. In low signal-to-noise ratio (SNR) environments, the hard-alignment techniques exhibit a high probability of beam misalignment.

Dual-polarized antenna systems, discussed in~\cite{Ref_Oes04,Ref_Asp07, Ref_Jia07, Ref_Col08, Ref_Kim10, Ref_Cho12}, are expected to be incorporated with mmWave systems. 
From a signal processing point-of-view, the orthogonal polarizations relax the Welch bound conditions that constrain the channel sounding time. Each desired polarization can be decoupled independently by using {reference signal (RS) sequences} which are designed to support high rank transmission in systems such as LTE and LTE-Advanced \cite{Ref_Nam12,Ref_Ses11, Ref_3GPP}. Numerous advantages exist, including the ability to sound the orthogonal polarizations in parallel for channel training, {multiplex information across the polarizations, and obtain a high level of immunity} to polarization mismatch between the base station and mobile users. Furthermore, a dual-polarized system may be more space efficient, allowing to have more antennas in a small form factor. To the best of our knowledge, outdoor mmWave systems over dual-polarized channels have not been reported yet except in our previous work~\cite{Ref_Son13, Ref_Son13_2}.

For connecting small cells and mobile users, we consider a mmWave system employing dual-polarized antennas. To align the beam direction of a large number of antennas, we develop a beam alignment algorithm which operates well even in the low SNR regime. First, with a simplified, dual-polarized channel model, an optimal set of combining and beamforming vectors at baseband and an analog beamforming vector under perfect channel state information (CSI) are discussed in order to provide criteria for practical combining and beamforming solutions. Then, we propose a soft-decision beam alignment (soft-alignment) algorithm that exploits the dual-polarized channel. Based on maximum likelihood estimation (MLE), the beam alignment algorithm estimates the channel subspace given our channel model assumptions, which consist of a sub-channel vector and a block matrix corresponding to propagation between two polarizations.  One stage of sampling is performed, followed by a post-processing stage consisting of two rounds of beam alignment. Based on the rough beam direction of the sub-channel vector estimated in the first round, a more accurate beam direction is estimated in the second round.

Our second contribution is an adaptive algorithm which selects the sounding time efficiently. We derive an approximate probability of beam misalignment based on a beam pattern analysis of the codebook used for beam alignment. With this approximate probability, the system is able to choose an efficient number of channel samples in order to ensure low probability of beam misalignment. Based on the formulation, the system changes the appropriate sounding time adaptively according to the channel environment to satisfy the predefined criterion.
Despite the severe path loss at mmWave frequencies, the proposed algorithm effectively scans the channel subspace with the minimum necessary number of sounding samples. 

In Section II, we describe a mmWave system employing dual-polarized antennas. In Section III, a practical soft-alignment algorithm is proposed for dual-polarized mmWave systems. In Section IV, an adaptive sounding algorithm is developed for the soft-alignment algorithm. In Section V, numerical results are presented to verify the performance of the proposed algorithms and Section VI details our conclusions.

Throughout this paper, $\mathbb{C}$ denotes the field of complex numbers, $\mathcal{CN}$ denotes the complex normal distribution, $\mathcal{N}$ denotes the normal distribution, {$\big[a,b\big]$ is the closed interval between $a$ and $b$, $\mathrm{U}\big(a,b\big)$ denotes the uniform distribution in the interval $\big[a,b\big]$,} $\mathbf{I}_{N}$ is the $N \times N$ identity matrix,  $\mathrm{E}[\cdot]$ is the expectation operator, $(\cdot)^*$ is the complex conjugate, $(\cdot)^H$ is the conjugate transpose, $\mathfrak{R}(\cdot)$ is the real part of complex number, $\mathfrak{I}(\cdot)$ is the imaginary part of complex number, $\mathbbm{1}_{[a,b)}$ is the indicator function,  $\| \cdot \|_p$ is the $p$-norm, $\odot$ is the Hadamard product, $\otimes$ is the Kronecker product, $\bA_{a,b}$, $\bA(:,a)$,  $\mathrm{vec}(\bA)$, $\big| \bA \big|$ denote $(a,b)^{th}$ entry, $a^{th}$ column vector, vectorization, and cardinality of the matrix $\bA$.


\section{System model}
\label{sec:System Model_1}

\subsection{System model}
\label{sec:System Model_2}
We consider a mmWave MIMO system {transmitting over block-fading channels} between a base station and mobile users. The  system employs dual-polarized antennas where each array of antennas is divided evenly into two groups, one of vertically polarized antennas and the other of horizontally polarized antennas. The base station has $M_{t}$ transmit antennas and two RF chains, one for each polarization. A mobile user has $M_r=2$ receive antennas\footnote{If we consider the large $M_r$ case, beam alignment at the receiver side should also be considered as in \cite{Ref_Hur13,Ref_Hur11,Ref_Son13,Ref_Son13_2}. Conducting beam alignment at both sides complicates our analyses. In this work, we consider a multiple-input single-output (MISO) channel for each polarization, {i.e., a single receive antenna for each of the} vertical and horizontal receive antenna groups.} with an RF chain for each polarization. An overview of the  mmWave system is shown in Fig. 1.

The input-output expression is represented by
\begin{align}
\label{eq:01}
y=\bz^H \big( \sqrt{\rho}\bH \bff s + \bn\big),
\end{align}
where $y$ is the received baseband signal, $\bz \in \mathbb{C}^{M_r}$ is the unit norm receive combining vector, $\rho$ is the transmit SNR, $\bH \in \mathbb{C}^{M_r \times M_t}$ is the block fading, dual-polarized channel matrix, $\bff \in \mathbb{C}^{M_t}$ is the unit norm transmit beamforming vector, $s \in \mathbb{C}$ is the data symbol subject to the power constraint $\mathrm{E}[|s|^2]\leq 1$ and $\bn=[n_{v}~n_{h}]^{T} \in \mathbb{C}^{M_r} \sim \mathcal{CN}(\mathbf{0},\bI_{M_r}) $ is the noise vector with independent and identically distributed (i.i.d.) entries for each polarization.
\begin{figure}
\centering
\includegraphics[width=0.493\textwidth]{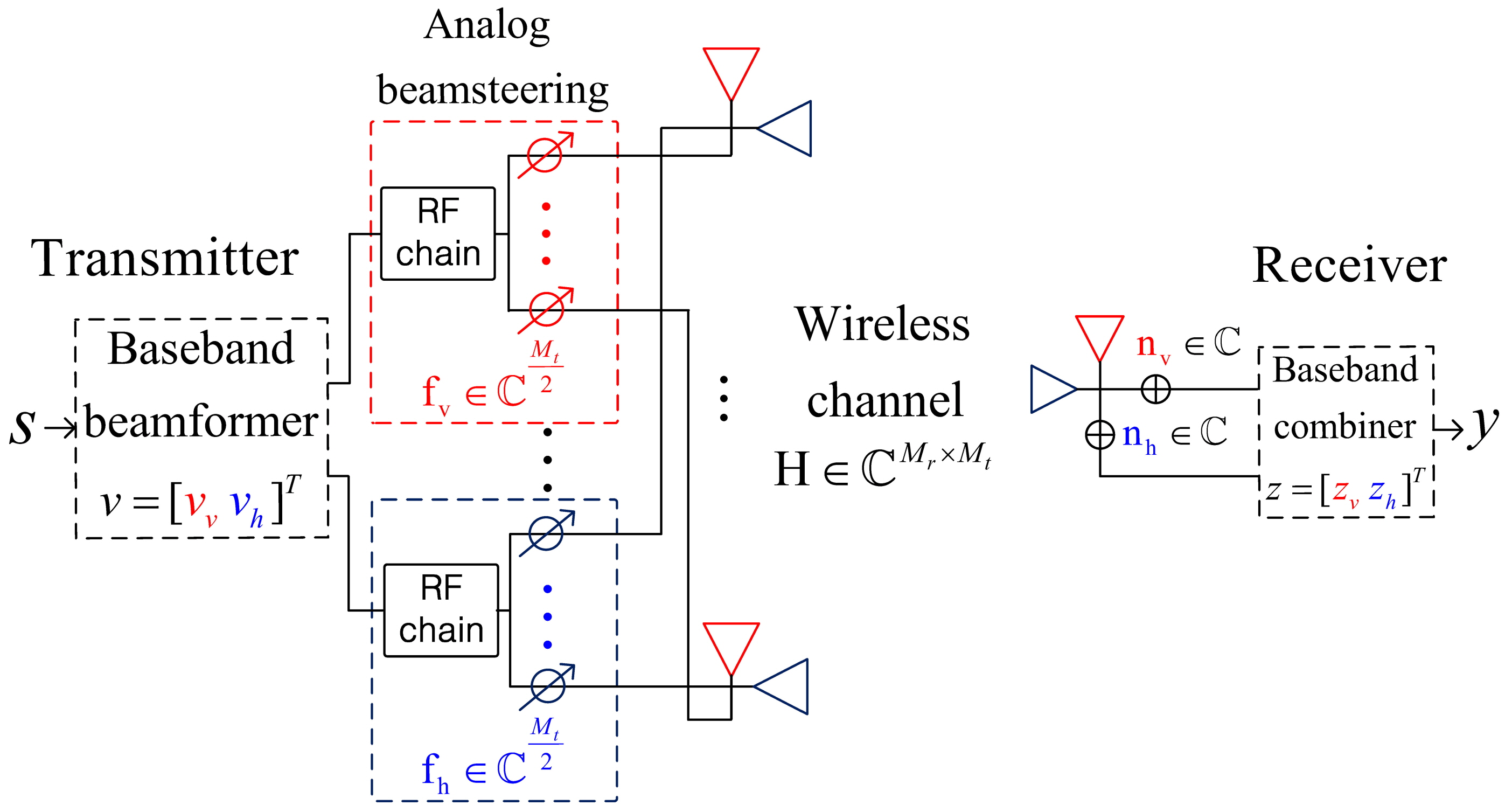}
\caption{An overview of the mmWave system.}
\end{figure}

As will become clear when the dual-polarized channel is examined, the transmit beamforming vector \(\bff\) is logically split into two unit-norm vectors corresponding to the analog beamforming weights. Let $\bff_{v} \in \mathbb{C}^{\frac{M_t}{2}}$ and $\bff_{h} \in \mathbb{C}^{\frac{M_t}{2}}$ respectively be the vertical and horizontal analog beamforming vectors.  The complete beamforming vector is then $\bff=[v_{v}\bff_{v}^T~v_{h}\bff_{h}^T]^T$, where $\bv=[v_{v}~v_{h}]^T\in\mathbb{C}^2$ is a unit norm weight vector that performs beamforming at baseband, such as maximum ratio transmission. The beam combining vector $\bz=[z_{v}~ z_{h}]^T \in \mathbb{C}^{2}$ accounts for the ability to perform arbitrary combining at baseband, such as maximum ratio combining.

At the base station, the analog beamforming vectors are constrained to a subset of \(\mathbb{C}^{\frac{M_t}{2}}\) and the antennas themselves are assumed to be arranged as a ULA.  {The set of vertical antennas forms a ULA with uniform element spacing,} and the horizontal set is similarly arranged.  The ability to control the gain and phase of each antenna at baseband is impractical due to the high cost and power consumption of many individual RF chains~\cite{Ref_Hur13,Ref_Hur11,Ref_IEEE802.11,Ref_Son13,Ref_Son13_2}. Instead, analog beamforming is performed with RF phase shifters and no gain control~\cite{Ref_Hur13}.  The constrained set of possible equal gain beamforming vectors in the {$N$-dimensional} complex space is
\begin{align}
\label{eq:02}
  \cB_{N} = \{\bw \in \mathbb{C}^{N} : (\bw \bw^H)_{\ell,\ell} = {1}/{N}, 1\le \ell \le N \}.
\end{align}

\subsection{Dual-polarized channel model}
\label{sec:04}
The mmWave channel differs from the conventional Rayleigh channel model, which assumes a rich scattering channel environment, and development of an appropriate model is necessary \cite{Ref_Tho14,Ref_Zha10,Ref_Rap12}. A general view of the dual-polarized mmWave channel follows from partitioning the channel into a block matrix form as
\begin{align}
\mathbf{H}&= \begin{bmatrix} \mathbf{h}_{vv}^H & \mathbf{h}_{vh}^H  \\ \mathbf{h}_{hv}^H  & \mathbf{h}_{hh}^H  \end{bmatrix}
\label{eq:03}
\end{align}
where $\bh_{ab} \in \mathbb{C}^{\frac{M_t}{2} \times \frac{M_r}{2}}$ is a sub-channel vector from polarization $b$ to $a$.

\begin{figure}[!t]
\centering
\subfloat[Dual-polarized MIMO scenario]{\includegraphics[width=0.244\textwidth]{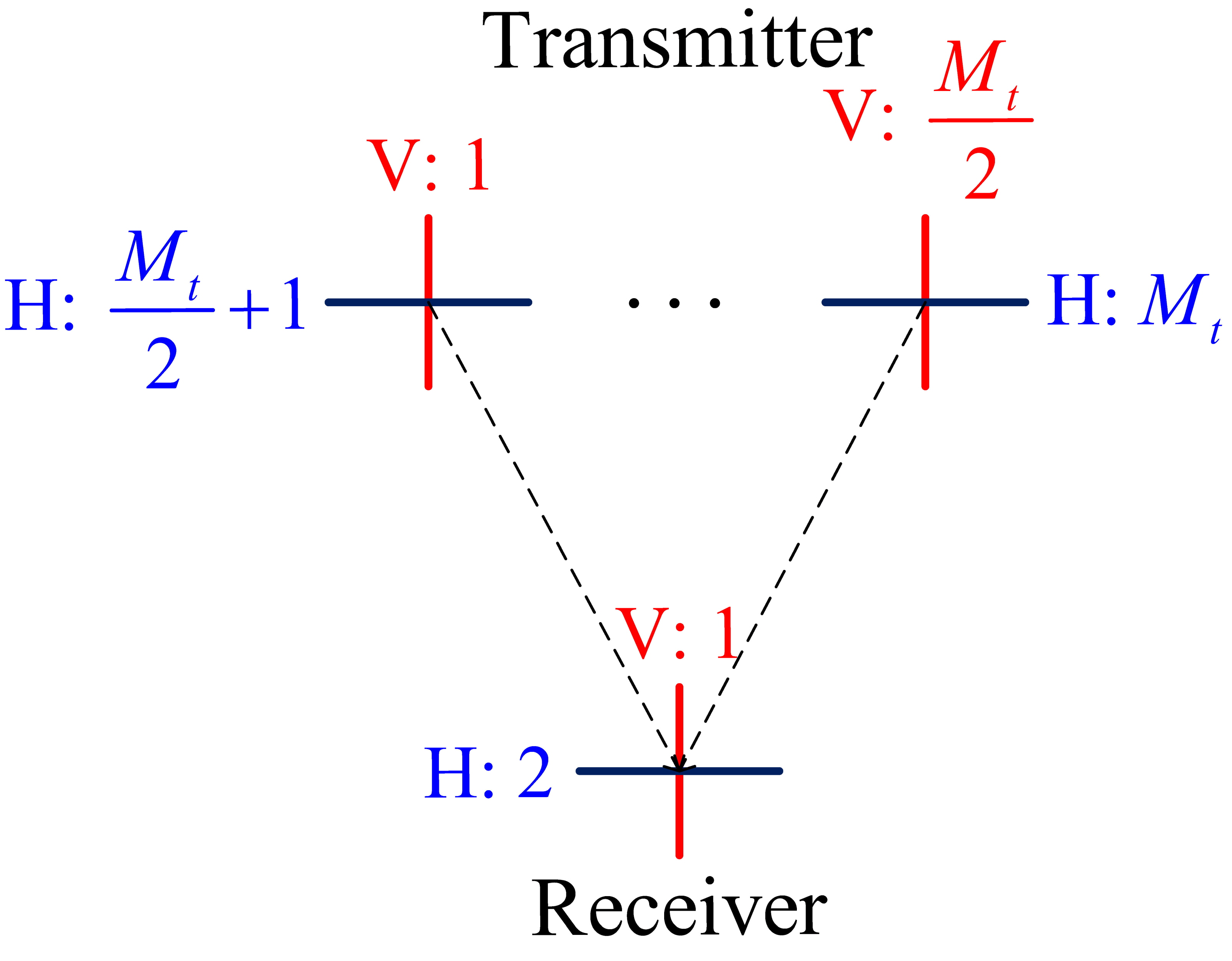}%
\label{fig_first_case}}
\hfil
\subfloat[Rotated-dual-polarized MIMO scenario]{\includegraphics[width=0.244\textwidth]{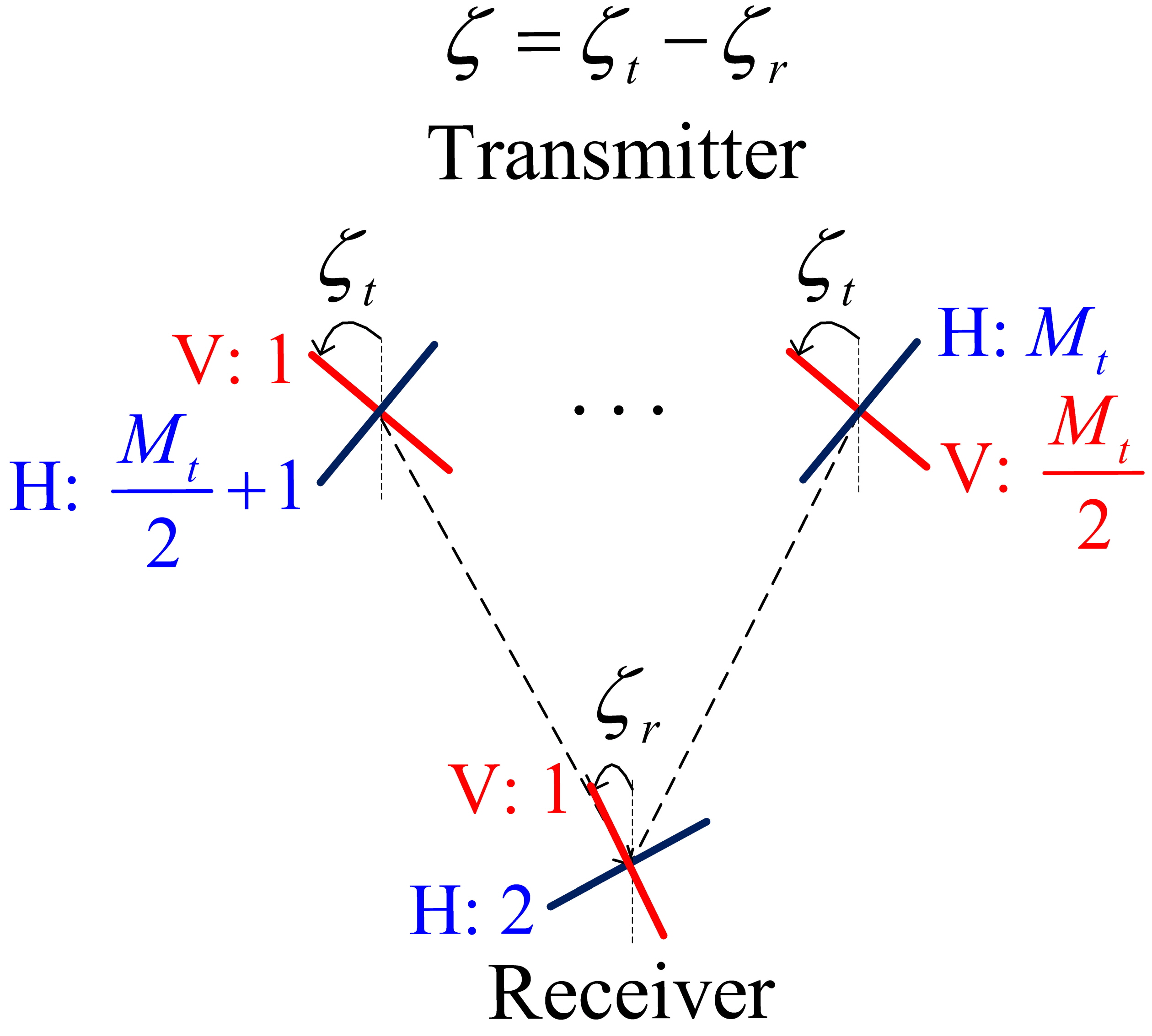}%
\label{fig_second_case}}
\caption{Rotated dual-polarized MIMO channel.}
\label{fig_sim}
\end{figure}

A dual-polarized mmWave channel model is proposed in~\cite{Ref_Son13}. However, this realistic channel model complicates analysis and hence a simpler channel model is needed. ~{Due to the dominant single path observed in most likely mmWave deployments \cite{Ref_Zha10,Ref_Rap12,Ref_Say13,Ref_Muh10}}, we approximate the mmWave channel with a single path whose angle of departure (AoD) from the base station to the mobile user is \(\theta\). Under the assumption that propagation between the different polarization pairs only varies by a scalar, the block matrix representation in (\ref{eq:03}) may be written as
\begin{align}
\bH &\simeq \bA \otimes \bh^H,
\label{eq:04}
\end{align}
where {$\bA \in \mathbb{C}^{M_r \times 2}$ is the polarization matrix} and $\mathbf{h}=\mathbf{a}_{\frac{M_{t}}{2}}({\theta})$ is the array response vector for the {single path with AoD \(\theta  \sim \mathrm{U}\big(\theta_{LB}, \theta_{UB}\big)\)}. The array response vector for a ULA is
\begin{align}
\label{eq:07}
\mathcal{A}_N&=\{ \ba_{N}(\theta)\in\mathbb{C}^N : \theta_{LB} \leq \theta \leq \theta_{UB} \}
\\
\label{eq:06}
\mathbf{a}_{N}({\theta})&=[1~e^{j\frac{2\pi d}{\lambda}\sin{{\theta}}}~\cdots~ e^{j\frac{2\pi d}{\lambda}(N-1)\sin{{\theta}}}]^{T} ,
\end{align}
where $d$ is the antenna element spacing and $\lambda$ is the wavelength.

In (\ref{eq:04}), the polarization matrix $\bA$ contains the gains and phase shifts across polarizations as
\begin{align*}
\bA= \begin{bmatrix} \alpha_{vv} & \alpha_{vh} \\ \alpha_{hv} & \alpha_{hh}  \end{bmatrix}= \begin{bmatrix}  \beta_{vv} \cos{\zeta} & -  \beta_{vh} \sin{\zeta}  \\    \beta_{hv}\sin{\zeta}  & \beta_{hh}  \cos{\zeta} \end{bmatrix} ,
\end{align*}
where $\beta_{ab} \in \mathbb{C}$ is the complex gain of the ray from polarization $b$ to $a$, and $\zeta$ is the difference in orientation between the base station and mobile users based on Malus' law~\cite{Ref_Jia07,Ref_Cho12} as shown in Fig. 2. We develop the sounding schemes based on the simplified channel model in (\ref{eq:04}), while the realistic channel model in~\cite{Ref_Son13} is used for numerical simulations presented in Section~\ref{Simulation Results}.

\subsection{Channel sounding for beam alignment}

The system considered in this paper sounds the channel \(L\) times {per coherence time block} with the training beamformers \(\bc[\ell]\in\mathbb{C}^{\frac{M_t}{2}},~ \left\| \bc[\ell] \right\|_2^2=\frac{1}{2}\) on each polarized antenna group.  Denote the set of beamformers used for sounding as \(\cC = \big\{ \bc[1],\cdots,\bc[L]\big\} \) and let the matrix of beamformers be \(\bC=\big[ \bc[1] \cdots \bc[L] \big] \in\mathbb{C}^{\frac{M_t}{2}\times L}\). The time required to sound the channel should be much less than the coherence time of the channel in order to maximize the useful time for information transmission. {In the $\ell$-th channel use,} the $\ell$-th beamforming vector $\bc[\ell]$ is modulated by using unit norm, precoded pilot sequences $\bv_v$ and $\bv_h$ {to generate training signals for the vertical and horizontal antenna groups.} Under the simplified channel in~(\ref{eq:04}), the received training signals on vertical and horizontal antenna groups before receive side combining are
\begin{align*}
	\begin{bmatrix}  \by_{v}^{H} [\ell] \\  \by_{h}^{H}[\ell]  \end{bmatrix}
	= \sqrt{\rho}\left(\bA \otimes \bh^H\right) \left( \begin{bmatrix} \bv_v^{H}\\ \bv_h^{H}  \end{bmatrix} \otimes \bc[\ell] \right)+
	\begin{bmatrix} \bn_{v}^{H}[\ell] \\  \bn_{h}^{H}[\ell] \end{bmatrix},
\end{align*}
where $\bn_{a}[\ell]$ is the $\ell$-th noise sequence for receive polarization $a$ with i.i.d. noise entries $n_{a}$ in (\ref{eq:01}).

The precoded pilot sequences $\bv_v$ and $\bv_h$ can be constructed {using RSs} found in standards such as LTE and LTE-Advanced \cite{Ref_Nam12, Ref_Ses11, Ref_3GPP}. The receiver decouples the four pairs of polarizations by using a different RS for each polarization. Demodulation is facilitated by the orthogonality of the RS. For example, the popular Zadoff-Chu sequences could be used for the RSs \cite{Ref_3GPP}.

 The training sample from polarization $b$ at the transmitter is recovered by a filter matched to the RS sequence vector $\bv_b$. The decoupled output  for receive polarization \(a\) is
\begin{align}
\label{eq:08}
y_{ab}[\ell]&= \by^H_a[\ell]\bv_b=\sqrt{\rho}\alpha_{ab}\bh^H\bc[{\ell}]+n_{ab}[\ell].
\end{align}
Let the observation vector for a polarization pair \(ab\) be
\begin{align}
\nonumber
\by_{ab}&=\big[y_{ab}[1], \cdots ,y_{ab}[L]\big]^H
\\
\label{eq:09}
&=\sqrt{\rho}\alpha_{ab}^{*}\bC^H\bh+\bn_{ab} \in \mathbb{C}^{L} ,
\end{align}
where the noise vector \(\bn_{ab}\) has i.i.d.\ \(\mathcal{CN}(0,1)\) entries.

\section{beam alignment Algorithm}
\label{sec:Beam alignment algorithm}
{In order to make up for the deficit in  system link budget~\cite{Ref_Hur13}} due to various channel losses and low power amplification, a mmWave system {must} employ beamforming and combining to increase the effective SNR.  However, the method of choosing a beamformer and combiner is non-trivial when the channel is not directly observable.  
Reliable operation of previously proposed beamforming algorithms~\cite{Ref_Hur13,Ref_Hur11,Ref_IEEE802.11,Ref_Son13,Ref_Son13_2} is limited to an unnecessarily high SNR range. Links that would otherwise be possible with good beamforming gain cannot be formed.  The common factor that limits performance at low SNR is the hard-decision being made in the cited algorithms.

It is natural to consider a soft-decision algorithm to improve beamforming reliability and performance.  Instead of choosing one of the beamformers from the set used for channel sounding (referred to as a hard-decision), a soft-decision algorithm may choose any feasible beamformer. In the hard-decision algorithms, the probability of choosing the optimal beamformer is fundamentally limited by the SNR \(\rho\).  More observations lead to a higher probability of choosing a good beamformer, but the probability of choosing the {optimal beamformer in hard-alignment} is eventually limited by the  SNR and not the number of observations.  On the other hand, the performance of our soft-decision algorithm is not fundamentally limited by the SNR and hence can scale with the number of channel sounding observations.

\subsection{Full CSI beam alignment}
\label{sec:Full CSI beam alignment: Point-to-point single data stream}
Before investigating practical mmWave beamforming solutions, we pause to consider optimal beamforming vectors under perfect CSI conditions for the simplified channel model in (\ref{eq:04}). The full CSI beamforming solution gives an upper bound on performance for practical systems under the assumed system and channel model and it offers insight on the design of practical beamforming algorithms.

In order to maximize the achievable rate of the system, the magnitude of the effective channel gain should be maximized~\cite{Ref_Lov03} by choosing optimal beamforming and combining vectors. The maximizers of the effective gain are
\begin{align}
  \label{eq:10}
   (\check{\bz} ,\check{\bff}  )=\argmax_{{\bz} \in \mathbb{C}^2,{\bff} \in \mathbb{C}^{M_t}}|{\bz}^H\bH{\bff}|^2.
\end{align}
Keep in mind that  \(\bz\) and \(\bff\) are unit norm. The solution to (\ref{eq:10}) is well known to be the dominant left and right singular vectors of $\bH$. Taking a closer look, the singular value decomposition (SVD) of $\bH$ is
\begin{align*}
\bH &= (\mathbf{U}_{\bA}\mathbf{\Sigma}_{\bA} \mathbf{V}_{\bA}^H ) \otimes (\mathbf{U}_{\bh} \mathbf{\Sigma}_{\bh} \mathbf{V}_{\bh}^{H})
\end{align*}
where $\mathbf{U}_{\bA} \mathbf{\Sigma}_{\bA} \mathbf{V}_{\bA}^{H}$ is the SVD of $\mathbf{A}$, and $\mathbf{U}_{\bh} \mathbf{\Sigma}_{\bh} \mathbf{V}_{\bh}^{H}$ is the SVD of $\bh^H$. Considering the constraints on the baseband beamforming vector $\bff_{\bA}$, the analog beamforming vector $\bff_{\bh}$ and the channel model in (\ref{eq:04}), the constrained maximizers are
\begin{align*}
\big(\check{\bz}_{\bA},\check{\bff}_{\bA},\check{\bff}_{\bh} \big)
&=\argmax_{({{\bz}}_{\bA},{{\bff}}_{\bA}, {{\bff}}_{\bh}) \in  \mathbb{C}^2\times\mathbb{C}^2\times\cB_{\frac{M_t}{2}} } \left|({\bz}_{\bA}^H \bA {\bff}_{\bA}) \otimes  ({\bh}^H {\bff}_{\bh})\right|^2.
\end{align*}
The optimal solutions for \(\check{\bz}_\bA \) and \(\check{\bff}_\bA\) are the dominant left and right singular vectors of $\bA$, respectively, i.e., $\check{\bz}_{\bA}=\bU_{\bA}(:,1)$, $\check{\bff}_{\bA}=\bV_{\bA}(:,1))$, because the combining and beamforming vectors $\bz_{\bA}$ and $\bff_{\bA}$ at baseband have no equal gain constraint. On the other hand, \(\check{\bff}_\bh\) is restricted by the equal gain constraint. An optimal solution is $\check{\bff}_\bh=\frac{\bh}{\|\bh\|_2}$ since \( \bh \) is assumed to be an array manifold vector in (\ref{eq:06}), which has equal gain entries.

Optimal and feasible combining and beamsteering vectors in (\ref{eq:10}) are given by
\begin{align}
  \label{eq:11}
	\check{\bz} = \check{\bz}_\bA,\quad \check{\bff} = \check{\bff}_\bA \otimes \check{\bff}_\bh .
\end{align}
The unit norm weight vectors $\bz$ and $\bv$ that perform combining and beamforming at baseband and the analog beamforming vectors $\bff_v=\bff_h$ at RF phase shifters on both vertical and horizontal antenna groups, defined in the input-output expression of (\ref{eq:01}), correspond to $\check{\bz}_{\bA}$, $\check{\bff}_{\bA}$, and $\check{\bff}_{\bh}$ in (\ref{eq:11}), respectively.

The beamformer for a sub-channel vector in the array manifold set is better described as a beamsteering vector.  Furthermore, the algorithm of choosing a good beamsteering vector is better described as a beam alignment algorithm. When the beamformer is chosen from the array manifold set, the mainlobe of the gain pattern is being steered to maximize gain in a specific direction.  In effect, narrow beams are created that must be aligned to maximize the effective channel gain.

\subsection{Beam alignment with noisy channel sounding}
\label{sec:Beam alignment with Noisy Sounding}
We now present a beam alignment algorithm that seeks to maximize the effective channel gain without direct knowledge of the channel.  The general idea is to observe the baseband output responses \(\by_{ab}\) in (\ref{eq:09}) {resulting from a set of training vectors \(\cC = \big\{ \bc[1],\cdots,\bc[L]\big\} \) and then choose a beamformer in a finite set \(\cE = \big\{ \bee_1,\cdots,\bee_Q \big\} \) with $Q$ codewords} given the observations.\footnote{The choice of beamformer in \(\cE\) for beam alignment is based on the multilevel codebook in~\cite{Ref_Hur13,Ref_Hur11}.}


\subsubsection{Limits for hard-alignment}
\label{sec:Reviews on hard-alignment}
Previously proposed hard-alignment algorithms show good performance at high SNR but are  unreliable at low SNR~\cite{Ref_Hur11,Ref_Hur13,Ref_Son13,Ref_Son13_2}.  In a hard-decision algorithm, the chosen beamformer is restricted to be one of the beamformers used for channel sounding, i.e., \(\cE = \cC\).  The estimated beamforming vector \(\hat{\bff}\) is chosen to be the training vector \(\bc[\hat{\ell}]\in\cC\) with test sample $y_{ab}[\ell]$ in (\ref{eq:08}) as
\begin{align}
\label{eq:12}
	\hat{\ell}&=\argmax_{\ell\in\{1,\ldots,L\}} |\sqrt{\rho}\alpha_{ab}\bh^H\bc[{\ell}]+n_{ab}[\ell] |^2
\\
&\doteq \argmax_{\ell=\{1,\ldots,L\}} \big| m_{ab}^{h}[\ell] +n_{ab}^{h}[\ell] \big|^2,
\nonumber
\end{align}
where $m_{ab}^{h}[\ell]$ denotes {the scaled correlation} between $\bc[\ell]$ and the sub-channel vector $\bh$ to be estimated and $n_{ab}^{h}[\ell]\sim\mathcal{CN}(0,1)$ is additive random noise that hinders good beam alignment.

The variance in the estimation of \( \bh \) in (\ref{eq:12}) 
is related to the mean and variance of the test sample $y_{ab}[\ell]$, i.e., $\mathrm{E}[m_{ab}^{h}[\ell]]$ and $\mathrm{E}[|n_{ab}^{h}[\ell]|^2]=1$. For the best beam alignment, the variance of noise $n_{ab}^{h}[\ell]$ should be small in comparison with the power of $m_{ab}^{h}[\ell]$ which is given by
\begin{align}
\label{eq:13}
|m_{ab}^{h}[{\ell}]|^2 &=  \frac{\rho M_t}{4} | \alpha_{ab} |^2  \mathrm{G}[\ell]  \leq \frac{\rho M_t}{4}| \alpha_{ab} |^2,
\end{align}
where $\mathrm{G}[\ell] \doteq  | \tilde{\bh}^H\tilde{\bc}[{{\ell}}] |^2 \le 1$ is the beamforming gain between the normalized sub-channel $\tilde{\bh}=\frac{\bh}{\left\|\bh \right\|_2}$ and the normalized training vector $ \tilde{\bc}[{{\ell}}]=\frac{\bc[{{\ell}}]}{\| \bc[{{\ell}}] \|_2}$ using $ \| \bh\|_2^2=\frac{M_t}{2}$  and $\|\bc[{{\ell}}]\|_2^2=\frac{1}{2}$.  Notice that increasing the number of samples \(L\) does not increase the upper bound.  Thus, adaptively choosing \(L\) in a hard-alignment algorithm does not help at low SNR, with fixed \( M_t \).


\subsubsection{Maximum likelihood soft-alignment}
\label{sec:Maximum likelihood soft-alignment}
In the proposed soft-alignment algorithm for beam alignment, multiple sounding samples $\by_{ab}$ in (\ref{eq:09}) are considered together, while each sounding sample $y_{ab}[\ell]$ in (\ref{eq:08}) is considered separately in the hard-alignment algorithm. Specifically, the algorithm chooses the beamformer in a predefined codebook \(\cE\)  that maximizes the likelihood function of the received samples.  In the soft-alignment algorithm, training vectors and beamformers do not need to be in the same set, i.e., \(\cE \ne \cC \), and each set is designed independently. In contrast, hard-alignment requires \(\cE = \cC\). The number \(Q\) of candidate codewords in \(\cE\) may be much larger in size than the number of sounding samples \(L\), i.e., $Q \geq L$, whereas $Q = L$ in hard-alignment. If $Q$ is large, it is possible to utilize a large number of codewords generating narrow transmit beams, which can increase the beam alignment performance. Using a large set of candidate codewords that generate narrow transmit beams can increase the beam alignment performance.

In the proposed soft-alignment algorithm, the channel sounding and beam alignment are conducted in  separate stages.
 In the first stage, the algorithm performs subspace sampling using the training vectors in the set \(\cC\) and collects the sounding samples in the vectors \(\by_{ab}\) for each polarization pair as in (\ref{eq:09}).  In the second stage, the beamformer from the codebook \(\cE\) is chosen along with the baseband beamformer.

The training vectors in the set \(\cC\) are important but have not yet been discussed.  
For our soft-alignment algorithm, the set of training vectors $\cC$ is designed to meet the criterion of the Welch bound as in\cite{Ref_San10}, i.e., $\bC\bC^H=\frac{L}{M_t}\bI_{\frac{M_t}{2}},~L\ge\frac{M_t}{2}$. To satisfy the criterion, the discrete fourier transform training vectors are designed by using an array vector with uniform phase shift of $\frac{2\pi}{L}$ given by
\begin{align*}
	&\mathbf{c}[{\ell}]= \frac{1}{\sqrt{M_t}}\big[1~e^{j\frac{2\pi}{L}\ell}~\cdots~e^{j\frac{2\pi}{L}(\frac{M_t}{2}-1)\ell}\big]^{T}
\end{align*}
for $\ell=1,\dots,L$. Note that, in comparison with a single-polarized system, the criterion of the Welch bound with equality training sequences is relaxed from $L \geq M_t$ to $L \geq \frac{M_t}{2}$ due to the parallel operation of vertical and horizontal sounding.

In the second stage, the complex channel gains $\alpha_{ab}$ and the sub-channel vector $\bh$ in (\ref{eq:04}) are estimated by using the observation vectors $\by_{ab}$ in (\ref{eq:09}). Typically, in conventional cellular systems, a minimum mean square error (MMSE) estimate of the channel is computed given the observation vectors~\cite{Ref_San10}. However, using the MMSE estimate of the channel is limited due to little, or more likely no, control over the gain of each individual antenna in a mmWave system.

Our proposed soft-alignment algorithm maximizes the likelihood function of the observation vector instead of estimating the channel with an MMSE estimator. Using high Ricean ${K}$-factor and ray-like propagation assumptions \cite{Ref_Muh10,Ref_Zha10}, the sub-channel vector is an array response vector, i.e., ${\bh} \in \mathcal{A}_{\frac{M_t}{2}}$. The directional characteristic of mmWave links observed in \cite{Ref_Pi11,Ref_Say13,Ref_Rap12,Ref_Muh10}, limits the space of practical beamforming vectors that must be searched. This enables mmWave systems to effectively perform MLE.

Initially, we will derive the MLE for a single polarization pair \(ab\), which is given by
\begin{align*}
(\hat{\alpha}_{ab}, \hat{\bh})=\argmax_{\bar{\alpha}_{ab} \in \mathbb{C}, {\bar{\bh}} \in \mathbb{C}^{\frac{M_t}{2}}} f(\by_{ab} | \bar{\alpha}_{ab}, {\bar{\bh}} ),
\end{align*}
where $f(\by_{ab} | \bar{\alpha}_{ab}, {\bar{\bh}} )$ is  the probability distribution function (pdf) for $\by_{ab}$ defined as
\begin{align*}
	f(\by_{ab} | \bar{\alpha}_{ab}, {\bar{\bh}} )=\frac{1}{\pi^L}\exp{(-\| \by_{ab}^H-\sqrt{\rho} \bar{\alpha}_{ab} {\bar{\bh}}^H\bC \|_2^2)}.
\end{align*}
The maximization of the pdf is equivalent to the minimization
\begin{align}
\label{eq:14}
(\hat{\alpha}_{ab}, \hat{\bh})=\argmin_{\bar{\alpha}_{ab} \in \mathbb{C}, {\bar{\bh}} \in \mathbb{C}^{\frac{M_t}{2}}} \| \by_{ab}^H-\sqrt{\rho}\bar{\alpha}_{ab} {\bar{\bh}}^H\bC \|_2^2
\end{align}
which needs to be minimized over $\bar{\alpha}_{ab}$ and ${\bar{\bh}}$. {The $2$-norm of dummy variable $\bar{\bh}$ is defined as $\| \bar{\bh} \|_2=\sqrt{\frac{M_t}{2}}$ because the sub-channel vector, to be estimated, is an array response vector, i.e., ${\bh} \in \mathcal{A}_{\frac{M_t}{2}}$.}

First, we consider the channel gain $\alpha_{ab}$. To estimate $\alpha_{ab}$, we differentiate the objective function over $\bar{\alpha}_{ab}^{*}$ as
\begin{align*}
\frac{\partial}{\partial  \bar{\alpha}_{ab}^{*}} \| \by_{ab}^H-\sqrt{\rho}\bar{\alpha}_{ab} {\bar{\bh}}^H\bC \|_2^2=\rho\bar{\alpha}_{ab}\|\bC^H{\bar{\bh}}\|_2^2-\sqrt{\rho}\by_{ab}^H\bC^H{\bar{\bh}}.
\end{align*}
{Then, the channel gain MLE is derived as
\begin{align}
\label{eq:15}
&\hat{\alpha}_{ab}=\frac{\by_{ab}^H\bC^H{\bar{\bh}}}{\sqrt{\rho}\|\bC^H{\bar{\bh}}\|_2^2}={\alpha}_{ab}  \frac{ \tilde{\bh}^H  \bar{\bh}}{\| \bar{\bh} \|_2}  +\sqrt{\frac{2}{\rho L}}n_{ab},
\end{align}
where $\tilde{\bh} \doteq \frac{\bh}{\left\|\bh \right\|_2}$ is the normalized sub-channel vector and $n_{ab} \doteq  \frac{\bn_{ab}^H (\bC^H\bar{\bh})}{\| \bC^H\bar{\bh} \|_2}$ follows $\mathcal{CN}(0,1)$.} Note that the effective channel $ \frac{\tilde{\bh}^H \bar{\bh}}{\| \bar{\bh} \|_2}$ influences the accuracy of the channel gain estimator.  Moreover, the variance of noise component decreases as \(L\) increases.


By plugging in the estimated channel gain $\hat{\alpha}_{ab}$, 
the MLE for the sub-channel vector becomes
\begin{align}
\hat{\bh}=\argmax_{{\bar{\bh}} \in \mathbb{C}^{\frac{M_t}{2}}} {\frac{ | \by_{ab}^H\bC^H{\bar{\bh}} |^2 }{\|\bC^H{\bar{\bh}}\|_2^2}} .
\label{eq:16}
\end{align}

The size of the feasible space for the MLE in (\ref{eq:16}) can be significantly reduced from $\mathbb{C}^{\frac{M_t}{2}}$ because the sub-channel vector is simply an array response vector under the high Ricean ${K}$-factor \cite{Ref_Muh10,Ref_Zha10}. This assumption simplifies the maximization of the likelihood function by not requiring a search over the large set of all feasible channel vectors.  Instead, a practically sized set of  equal-gain beamsteering vectors \(\cE= \{\bee_1,\dots,\bee_Q\}\subset\cB_{\frac{M_t}{2}}\) may be used as the feasible set for the maximization. With the feasible set \(\cE\), the estimated channel vector in (\ref{eq:16}) is \(\hat{\bh}=\sqrt{\frac{M_t}{2}} \bee_{\hat{q}}\) where
\begin{align}
\hat{q}&=\argmax_{  {q} \in \{1,\dots,Q \} } {  \frac{ | \by_{ab}^H \bC^{H} \bee_{q} |^2}{ \|  \bC^{H} \bee_{q} \|_2^2}}
\doteq \argmax_{  {q} \in \{1,\cdots,Q \} }| t_{ab}[q] |^2.
\label{eq:17}
\end{align}
In (\ref{eq:17}), test sample $t_{ab}[q]$ is divided into two components
\begin{align*}
t_{ab}[q] &= \frac{\sqrt{\rho}\alpha_{ab}\bh^{H}\bC\bC^H\bee_q}{\|\bC^H\bee_q \|_2}+\frac{\bn_{ab}^H\bC^H\bee_q}{\|\bC^H\bee_q \|_2} \\
&= {\sqrt{\frac{\rho  L}{2}}\alpha_{ab} \tilde{\bh}^{H}\bee_q}+\frac{\bn_{ab}^H\bC^H\bee_q}{\|\bC^H\bee_q \|_2}
\\
&\doteq  m_{ab}^{s}[q] + n_{ab}^{s}[q].
\end{align*}
The mean and variance of $t_{ab}[q]$ are $\mathrm{E}[m_{ab}^{s}[q]]$ and $\mathrm{E}[|n_{ab}^{s}[q]|^2]=1$. The power of $m_{ab}^{s}[q]$ is
\begin{align}
 { |m_{ab}^{s}[q]|^2} &=\frac{\rho L}{2} | \alpha_{ab} |^2 \mathrm{G}[{q}]  \le \frac{\rho L}{2} | \alpha_{ab} |^2,
\label{eq:18}
\end{align}
where $\mathrm{G}[q] \doteq \big| \tilde{\bh }^H {\bee}_{{q}}\big|^2 \le 1$ is the beamforming gain 
between the normalized sub-channel and the codeword vector.

At this point, it is interesting to compare the maximization in (\ref{eq:17}) to the hard-decision maximization in (\ref{eq:12}). If \(L={\frac{M_t}{2}}\), the upper bound for the power of $m_{ab}^{s}[q]$ in (\ref{eq:18}) is the same as for the hard-decision algorithms in (\ref{eq:13}).  However, \(L\) is a system parameter that
may be varied to control the upper bound in (\ref{eq:18}). 
A method of efficiently selecting $L$ will be developed in Section \ref{sec:Adaptive sounding algorithm for a beam alignment}.


\subsubsection{Optimal combining for MLE}
{Until now, the beam alignment algorithm has only considered a single observation vector $\by_{ab} \in \mathbb{C}^{L}$. The estimation of \(\bh\) may be improved by linearly combining the observation vectors \(\by_{ab}\),
\begin{align*}
&\begin{bmatrix} \mathbf{y}_{vv}^H & \mathbf{y}_{vh}^H  \\ \mathbf{y}_{hv}^H  & \mathbf{y}_{hh}^H  \end{bmatrix}=\sqrt{\rho} \big(\bA \otimes \bh^H\bC \big) +\begin{bmatrix} \mathbf{n}_{vv}^H & \mathbf{n}_{vh}^H  \\ \mathbf{n}_{hv}^H  & \mathbf{n}_{hh}^H  \end{bmatrix}.
\end{align*}
Let the linear combination of \(\by_{ab}\) and some unit-norm vectors \(\bar{\bz}, \bar{\bff} \in \mathbb{C}^2 \) give the combined MLE for \(\bh\) defined as
\begin{align*}
(\hat{\bz},\hat{\bff},\hat{\bh})=\argmax_{({\bar{\bz}},{{\bar\bff}}, \bar{\bh}) \in  \mathbb{C}^2\times\mathbb{C}^2\times\mathbb{C}^{\frac{M_t}{2}} } {\frac{ \bigg| \bar{\bz}^H \begin{bmatrix} \mathbf{y}_{vv}^H & \mathbf{y}_{vh}^H  \\ \mathbf{y}_{hv}^H  & \mathbf{y}_{hh}^H  \end{bmatrix} (\bar{\bff} \otimes \bC^H{\bar{\bh}}) \bigg|^2 }{\|\bC^H{\bar{\bh}}\|_2^2}} .
\end{align*}
The question remains on how to choose \(\hat {\bz}\) and \(\hat {\bff}\).
Optimal combining is achieved when \( \hat{\bz} \) and \(\hat{\bff}\) are the dominant left and right singular vectors of $\bA$, respectively. 
However, the optimal combiners rely upon some knowledge of \(\bA\). If \(\bA\) is not known or cannot be accurately estimated, the performance of these combining schemes suffers  especially at low SNR.}

\subsubsection{Joint MLE}\label{sec:joint_MLE}
Instead, we develop the joint MLE for \(\bA\) and \(\bh\).  The likelihood function that will be maximized is
\begin{align*}
	f(\by_{vv},\by_{vh},\by_{hv},\by_{hh}| \bA,\bh ) =
	\prod_{a,b \in \{ v,h\}} f(\by_{ab}| \alpha_{ab},\bh ),
\end{align*}
which can be manipulated in a similar manner as in Section \ref{sec:Maximum likelihood soft-alignment} to give the MLE for \(\hat{\bh}=\sqrt{\frac{M_t}{2}} \bee_{\hat{q}}\), where \(\hat{q}\) is
\begin{align}
\label{eq:19}
\hat{q}&=\argmax_{  {q} \in \{1,\dots,Q \} } { \sum_{a,b \in \{ v,h\}} \frac{  |\by_{ab}^H \bC^{H} \bee_{q} |^2}{ \|  \bC^{H} \bee_{q} \|_2^2}}
\\
\nonumber
&= \argmax_{  {q} \in \{1,\cdots,Q \} }  \sum_{a,b \in \{ v,h\}} | t_{ab}[q] |^2.
\end{align}
A maximizer in (\ref{eq:19}) is found by searching through a codebook \(\cE\) as was done in (\ref{eq:17}). Finally, the joint MLE for the entries of \(\hat{\bA}\) is identical to (\ref{eq:15}) with \(\bar{\bh}=\hat{\bh}\).

\subsection{Extension to multi-round alignment}
\label{sec:Extension to multi-round alignment}
We fully elaborate our proposed soft-alignment algorithm based on the previous results.  After receiving sounding samples $\by_{ab} \in \mathbb{C}^{L}$ in (\ref{eq:09}), the proposed algorithm performs multiple rounds of soft-alignment using the same $L$ sounding samples.\footnote{In this work, the same \(L\) samples are used during all rounds of beam alignment at the receiver side, while in~\cite{Ref_Hur13,Ref_Hur11} a separate sampling is performed for each round. Thus, no extra time slot for sounding is necessary for multi-round beam alignment.} Each successive round of beam alignment refines the previous choice by using larger codebooks with narrower beamformers.\footnote{Having multiple rounds for beam alignment may increase the beam alignment performance. However, we only consider two rounds of beam alignment due to practical issues, e.g., small coherence times of the block fading channel and computational complexity at the receiver.} Note that multiple rounds of beam alignment only increases the computational complexity and not the overhead of channel sounding.

 In the first round of beam alignment, the system makes a preliminary decision for the beamforming vector corresponding to the sub-channel vector. The first level codebook \(\cE_1\) for the first round contains $Q_1=\frac{M_t}{2}$ orthogonal codewords which are designed to satisfy $\bee_{c}^H\bee_{d}=0,~c \ne d$. This assumption guarantees the likelihood samples are uncorrelated since $\mathrm{E}[|t_{ab}[c]^H t_{ab}[d]|^2]=\mathrm{E}[|\bee_{c}^H\bee_{d}|^2]=0$. Recall the assumption that the sub-channel vector $\bh$ is accurately described by an array response vector in (\ref{eq:06}). Since the beam-width of the channel vector is quite narrow for large \(M_t\), the channel vector might be orthogonal to all codewords except the properly estimated beam alignment codeword and its neighbor codewords. We address the range that is covered by the optimal codeword $\bee_{\check{q}}$ and its neighboring codewords $\bee_{\check{q} \pm 1}$ as a rough beam direction. {Note that the index for the codeword which is optimally aligned under perfect CSI condition of \(\bh \) is defined as
\begin{align*}
\check{q}=\argmax_{q \in \{1,\cdots,Q \}} \big| {\bh}^{H}\bee_q \big|^{2}.
\end{align*}
Beyond the rough beam direction,} we can approximate the power of the other likelihood samples as  $|{t}_{ab}[{q}]|^2\simeq |{n}_{ab}^{s}[q]|^{2}$ since ${\bh}^H\bee_q\simeq 0,~q \in \{ 1,\cdots,Q_1 \}  \backslash \{\check{q},\check{q}\pm 1\}$.

In later rounds of beam alignment, only beamformers which cover the range of the rough beam direction from the previous rounds are considered, allowing for more accurate beam alignment over many iterations. For example, the second level codebook \(\cE_2\) should be larger than the first, with each beamformer covering a smaller area. In this work, the number of codewords in \(\cE_2\) is set to $ Q_2=2M_t$. Only the set of codewords which cover the rough beam direction from the previous round are selected as the feasible set for beam alignment. With the estimated sub-channel vector $\hat{\bh}$ and the block matrix $\hat{\bA}$, mobile user computes an optimal combining and beamforming vector an identical way to the full CSI case in (\ref{eq:11}). Note that the beamforming vector can be fed back from the receiver to the transmitter in frequency division duplexing (FDD) systems. 

\section{Adapting sounding time for beam alignment}
\label{sec:Adaptive sounding algorithm for a beam alignment}

In the proposed soft-alignment algorithm, beam alignment performance is proportional to the sounding time $L$. Utilizing a large $L$ guarantees good beam alignment performance. However, a large number of sounding samples imposes a heavy burden on the overhead of the system, especially in the case of a fast fading channel. In addition, the beam alignment performance also varies depending on channel conditions, e.g., SNR $\rho$. To handle this trade-off problem, the base station needs to adaptively select {$L$ based on the SNR $\rho$.} In this section, an approximate probability of beam misalignment is derived to aid in choosing \(L\). Based on the error probability, we propose an adaptive sounding algorithm which adjusts $L$ according to the channel environment.

\subsection{Probability of beam misalignment}
\label{sec:Bnalysis on the probability of beam misalignment}

In the proposed  algorithm, a more accurate beam direction is estimated in the second round based on the rough beam direction estimated in the first round of beam alignment
, as defined in Section \ref{sec:Extension to multi-round alignment}. When the estimated rough beam direction does not contain the beam-width covered by the optimal codeword, the system fails to align the beamformer to the channel subspace. Thus, we define the event $\hat{q}  \not\in \{ \check{q},\check{q} \pm 1\}$ as the beam misalignment event.

The probability of beam misalignment is defined as
\begin{align*}
\mathrm{P}_{\mathrm{mis}}& \doteq \mathrm{Pr}\big( \{\hat{q}  \not\in \{ \check{q},\check{q} \pm 1\} \}  \big)
\\
&=\mathrm{Pr}\big( \max_{\check{q},\check{q} \pm 1} |  {t}[q]  |^{2} \leq   \max_{q \in \{ 1,\cdots,Q_1 \} \backslash \{\check{q},\check{q}\pm 1\}}   |   {t}[q]  |^{2} \big)
\\
& \leq \mathrm{Pr}\big(  |  {t}[\check{q}]  |^{2} \leq   \max_{q \in \{ 1,\cdots,Q_1 \} \backslash \{\check{q},\check{q}\pm 1\}}   |   {t}[q] |^{2} \big)
\\
&\doteq \mathrm{P}_{\mathrm{mis}}^{\mathrm{up}},
\end{align*}
where the upper bound $\mathrm{P}_{\mathrm{mis}}^{\mathrm{up}}$ follows from $\max_{q=\check{q},\check{q} \pm 1} |  {t}[q] |^{2} \geq  | {t}[\check{q}]  |^{2}$. Note that in this section, the polarization index $ab$ and the index $s$, which denotes the soft-alignment algorithm, are dropped for simplicity.

As discussed in Section \ref{sec:Extension to multi-round alignment}, we assume 
$| {t}[{q}] |^{2} \simeq |{n}[q]|^{2},~q \in \{ 1,\cdots,Q_1 \} \backslash \{\check{q},\check{q}\pm 1\}$. Under this assumption, 
$\mathrm{P}_{\mathrm{mis}}^{\mathrm{up}}$ is approximated as
\begin{align}
\label{eq:20}
\mathrm{P}_{\mathrm{mis}}^{\mathrm{up}} &\simeq \mathrm{Pr}( \bX - \bZ \leq 0 ),
\\
\nonumber
\bX &\doteq  |  {m}[\check{q}] +  {n}[\check{q}] |^{2},
\\
\nonumber
\bZ &\doteq \max_{q \in \{ 1,\cdots,Q_1 \} \backslash \{ \check{q},\check{q}\pm 1\}}   |   {n}[q] |^{2}.
\end{align}
In (\ref{eq:20}), $\bX$ is the power of the optimal likelihood sample and $\bZ$ is the maximum noise power among $Q_1-3$ noise samples.


First, the cumulative distribution function (cdf) of $\bX$ can be approximated as the noncentral chi-squared distribution with two degrees of freedom, i.e.,
\begin{align}
\label{eq:21}
\mathrm{F}_{\bX}(x)&\simeq\Big( 1-Q_{1}\big(\sqrt{2\mu_x^2},\sqrt{2x}\big) \Big)\mathbbm{1}_{[0,\infty)}(x),
\end{align}
where $\mu_x^{2} = \mathrm{E}\big[|m[\check{q}]|^2\big]$ is the noncentrality parameter of $\bX$ and $Q_{1}( \cdot , \cdot  )$ is the Marcum Q-function \cite{Ref_Pro01}. The approximated cdf of $\bX$ is derived in Appendix \ref{sec:A}.

The power of each noise sample $\bN\doteq | {n}[q] |^2$ follows the central chi-squared distribution with two degrees of freedom \cite{Ref_Pro01}. The cdf of $\bN$ is defined as
\begin{align*}
\mathrm{F}_{\bN}(n)=\big(1-e^{-n}\big)\mathbbm{1}_{[0,\infty)}(n).
\end{align*}
Then, the cdf of $\bZ$ is derived with the binomial series expansion,
\begin{align*}
\mathrm{F}_{\bZ}(z)&=\big(\mathrm{F}_{\bN}(z)\big)^{(Q_1-3)} =\sum_{q=0}^{Q_1-3}\binom{Q_1-3}q(-1)^qe^{-qz}.
\end{align*}

Based on the distribution of $\bX$ and $\bZ$, an upper bound on the probability of beam misalignment in (\ref{eq:20}) is derived as
\begin{align}
\nonumber
&{\mathrm{P}_{\mathrm{mis}}^{\mathrm{up}}(\mu_x^2,Q_1)}
\\
\nonumber
& \doteq 1- \mathrm{Pr}( \bX - \bZ > 0 )
\\
\nonumber
&=1 - \int_{0}^{\infty}  \bigg( \int_{0}^{x}f_{\bZ}(z) dz \bigg)  f_{\bX}(x)  dx
\\
\nonumber
&=1- \int_{0}^{\infty} \bigg(  \sum_{q=0}^{Q_1-3} \binom{Q_1-3}{q} (-1)^{q}e^{-qx} \bigg) f_{\bX}(x)    dx
\\
\nonumber
&=1- \sum_{q=0}^{Q_1-3} \binom{Q_1-3}{q} (-1)^{q} \int_{0}^{\infty}  e^{-qx}  f_{\bX}(x)  dx
\\
\nonumber
&=\sum_{q=1}^{Q_1-3}  \binom{Q_1-3}{q} (-1)^{q+1}  \mathrm{M}_{\bX}(-q)
\\
\label{eq:22}
&=\sum_{q=1}^{Q_1-3}  \binom{Q_1-3}{q}\frac{(-1)^{q+1}\exp\left(\frac{-\mu_x^2q}{1+q}\right)}{1+q},
\end{align}
where 
$\mathrm{M}_{\bX}(t)=\mathrm{E}\left[e^{t\bX}\right]= \frac{\exp{\frac{\mu_x^2 t}{1-t}}}{1-t}$ for $t<1$ is the moment-generating function in \cite{Ref_Pro01}.

\subsection{Adaptive sounding algorithm for soft-alignment algorithm}
In this work, the sounding time $L$ for the proposed soft-alignment algorithm is adapted according to the  probability of beam misalignment in (\ref{eq:22}). To develop an adaptive sounding algorithm, the noncentrality parameter $\mu_x^{2}$ in (\ref{eq:21}) for the probability of beam misalignment should be derived first. The noncentrality parameter $\mu_x^{2}$ is a function of the beamforming gain defined as
\begin{align}
\label{eq:23}
\mathrm{G}[{\check{q}}] =  | \tilde{\bh}^H \bee_{\check{q}} |^2.
\end{align}
We derive an upper bound for the expected beamforming gain $\mathrm{E}\big[\mathrm{G}[{\check{q}}] \big] < \varsigma^2$ in Lemma $2$ of Appendix \ref{sec:B}. Then, a tighter upper bound \(\nu\) for the noncentrality parameter $\mu_x^2$, which is a function of $\varsigma^2$, is defined by substituting the upper bound $\varsigma^2$ into $\mathrm{E}\big[\mathrm{G}[{\check{q}}] \big]$. In the proposed algorithm, we use this upper bound $\mu_x^2 \leq \nu$ as a noncentrality parameter instead of $\mu_{x}^{2}$.

With a predefined target error probability \(\epsilon\), the sounding time $L$ is adaptively adjusted as a function of the SNR $\rho$. \footnote{If system parameters $M_t$, $Q$, and $\epsilon$ are fixed, the adaptive sounding algorithm is a function of $\rho$. Note that SNR of mmWave channels highly depends on the path loss, which is not instantaneously changed, due to the high Ricean ${K}$-factor and poor scattering channel environments. Thus, the sounding time should not be adapted for every channel instant.}  Specifically, the sounding time should satisfy
{
\begin{align}
\label{eq:24}
\hat{L} = \argmax_{\ell \in\mathbb{N}} \mathrm{P}_{\mathrm{mis}}^{\mathrm{up}}(\nu,\ell) \text{ s.t. } \mathrm{P}_{\mathrm{mis}}^{\mathrm{up}}(\nu,\ell) < \epsilon
\end{align}
where \(\mathrm{P}_{\mathrm{mis}}^{\mathrm{up}}(\nu,\ell)\) is defined in (\ref{eq:22}) and $\nu$ is used as an upper bound of $\mu_x^{2}$ in \(\mathrm{P}_{\mathrm{mis}}^{\mathrm{up}}(\nu,\ell)\).}  The final adaptive sounding time \(L\) must satisfy the Welch bound with equality $(L \geq \frac{M_t}{2})$ in  \cite{Ref_San10} and is therefore chosen to be
\begin{align*}
L=\max \left(\hat{L},\frac{M_t}{2} \right).
\end{align*}


\section{Simulation results}
\label{Simulation Results}
In this section, we present numerical performance results for the proposed algorithm, which combines the soft-alignment estimation algorithm with the adaptive channel sounding.  We consider two performance metrics: 1) expected beamforming gain \(G_{BF}\) and 2) expected data rate \(R\).  The metrics are defined as
\begin{align}
G_{BF}&= \mathrm{E} \big[\big| \hat{\bz}^H \bH \hat{\bff} \big|^2/\big| \check{\bz}^H \bH \check{\bff} \big|^2 \big],
\\
R&=\mathrm{E} \big[ \log_2\big(1+\rho \big| \hat{\bz}^H \bH \hat{\bff} \big|^2\big)  \big] ,
\end{align}
where $\hat{\bz},\hat{\bff}$ are the estimated combining and beamforming vectors and $\check{\bz},\check{\bff}$ are the optimal combining and beamforming vectors.

The numerical results were obtained from Monte Carlo simulations of over $10000$ independent channel realizations with the following parameters. For the simulations, we adopt the realistic channel model in~\cite{Ref_Son13,Ref_Son13_2}. We consider a street geometry \cite{Ref_Zha10} with a line-of-sight path and three first order reflected paths from both the wall of buildings and the ground. We assume the upper and lower bounds of the range of AoD are $\theta_{UB}=-\theta_{LB}=\frac{\pi}{3}$ \cite{Ref_Hur13} and $\zeta \sim \mathrm{U}\big(0,\frac{\pi}{2}\big)$. The Ricean ${K}$-factor is set to $13.5~\mathrm{dB}$  based on the channel measurements in \cite{Ref_Muh10}.  The reciprocal of the cross polar discrimination value, which represents the ability to distinguish the polarization difference of the antennas~\cite{Ref_Cal07}, is set to $\chi=0.2$. We assume 6 bit phase control registers in the analog beamforming hardware.

\begin{figure}[!t]
\centering
\subfloat[$\rho=-4$ dB]{\includegraphics[width=0.2435\textwidth]{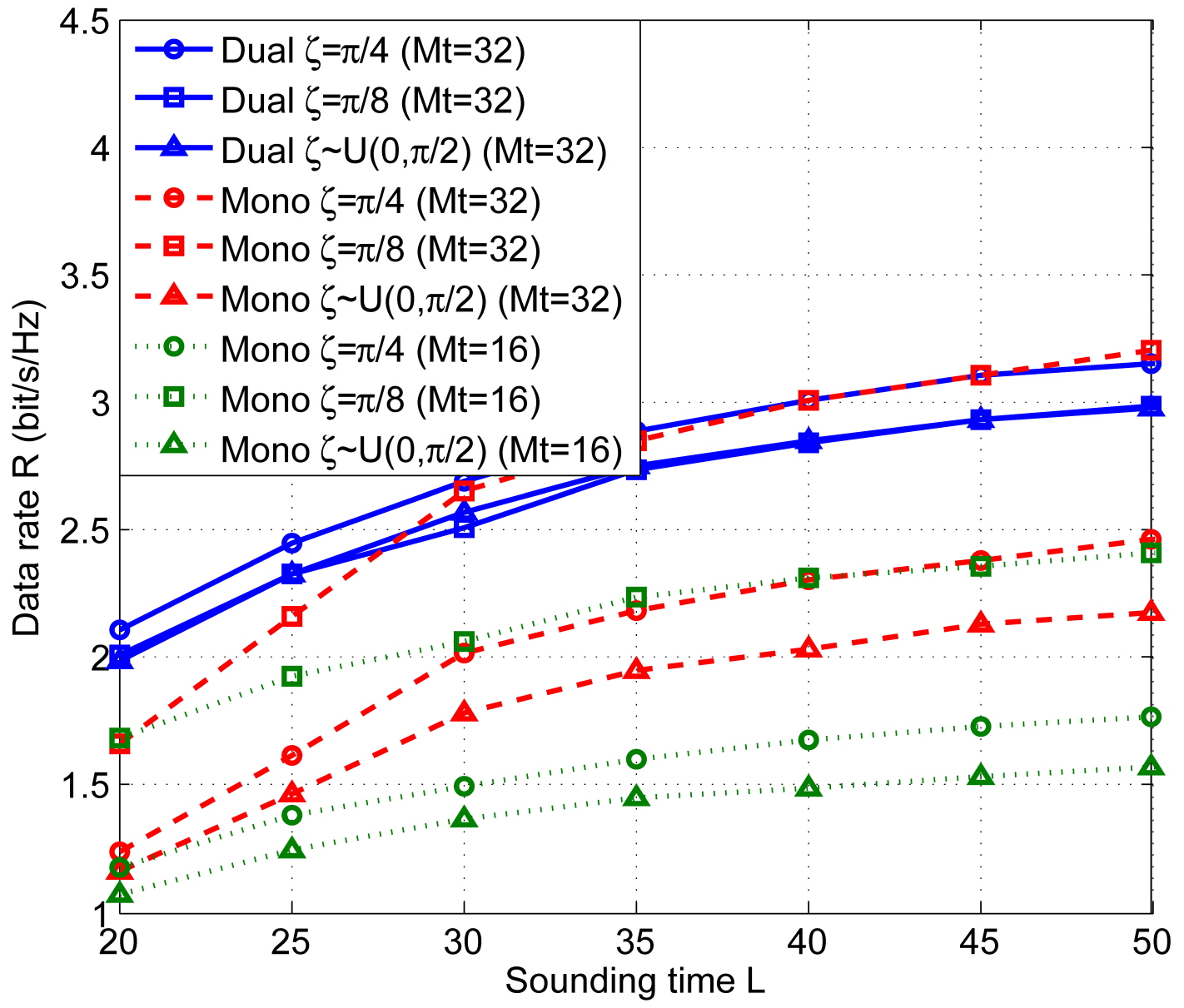}}
\hfil
\subfloat[$\rho=-6$ dB]{\includegraphics[width=0.2435\textwidth]{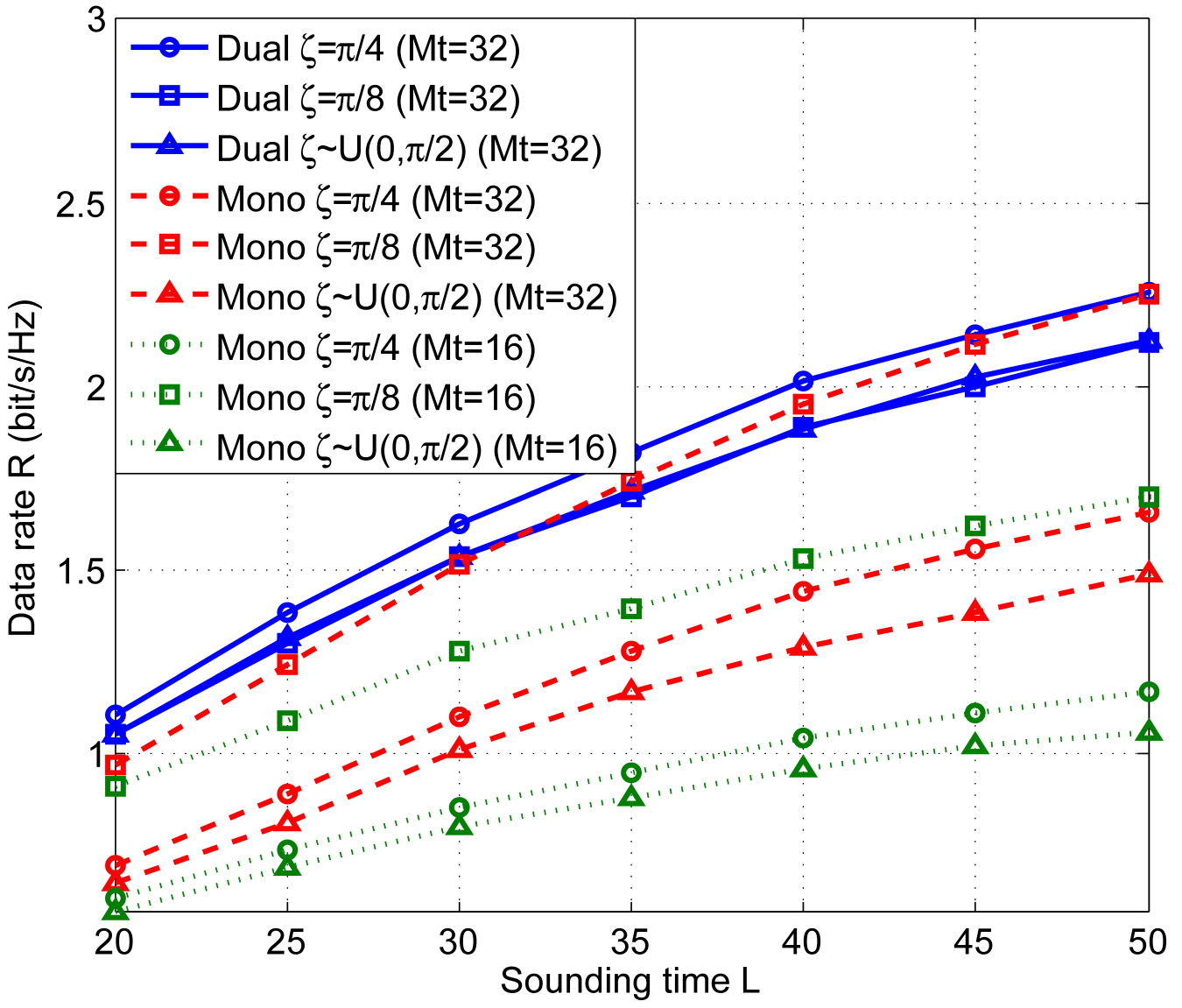}}
\hfil
\caption{Data rate of soft-alignment algorithm under both a mono-polarized channel and dual-polarized channel against sounding time $L$ (Mono polarized system: $M_t=16,~32$, $M_r=1$, Dual polarized system: $M_t=32$, $M_r=2$).}
\end{figure}

\begin{figure}[!t]
\centering
\subfloat[$M_t=32,~ M_r=2$]{\includegraphics[width=0.2435\textwidth]{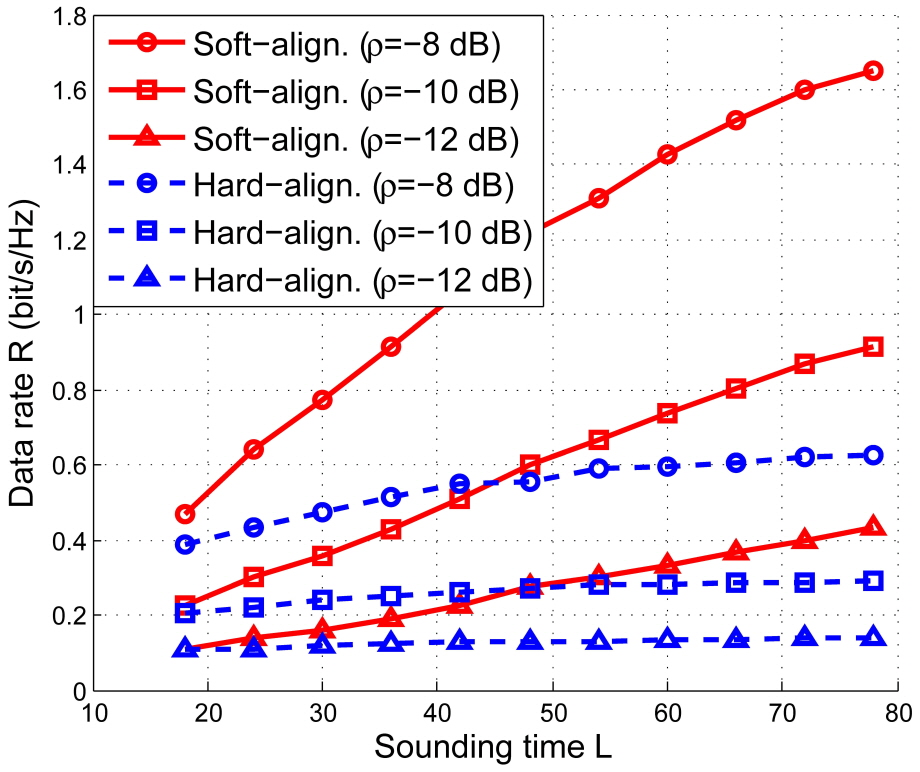}}
\hfil
\subfloat[$M_t=64,~ M_r=2$]{\includegraphics[width=0.2435\textwidth]{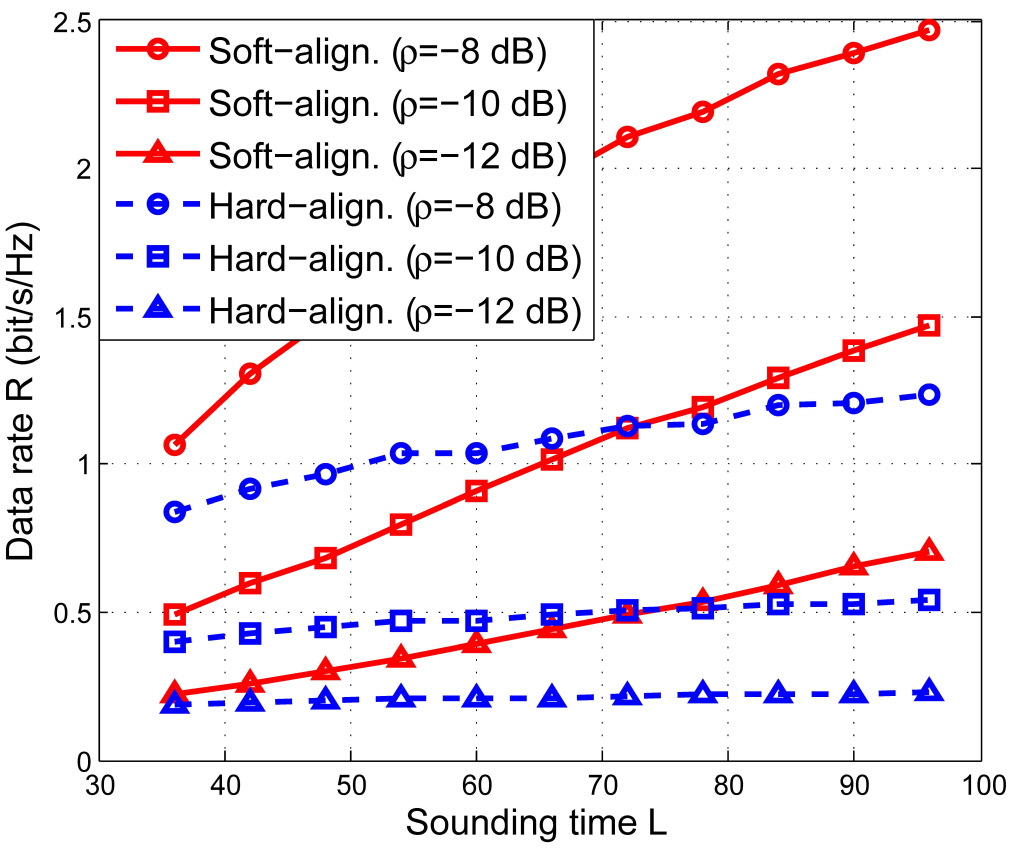}}
\caption{Data rate of soft and hard beam alignment algorithm against sounding time $L$ $(M_t=32,~64,$ $M_r=2,$ $\chi=0.2).$}
\end{figure}

In Figs. 3a and 3b, the data rate of the proposed soft-alignment algorithm for a dual-polarized system is compared with that of a mono-polarized {system as a function of the total sounding time $L$.} The simulation results are presented in different cases of polarization angle mismatch between the base station and mobile users, {i.e., $\zeta=\frac{\pi}{8},\frac{\pi}{4}$ and $\zeta \sim \mathrm{U}\big(0,\frac{\pi}{2}\big)$.} {In the first scenario of the system setup,} the number of transmit antennas for the mono-polarized system {is the same as that of the dual-polarized system.} {In the second scenario, the number of transmit antennas for the mono-polarized system is half that of the dual-polarized system, which follows from the assumption of a fixed area for antenna deployment.} This is possible because utilizing a dual-polarized array allows the more antennas to be packed into a restricted space. The data rate of the mono-polarized system approaches that of the dual-polarized system only under a special case when the polarization angle between the base station and mobile users is almost perfectly aligned, which would rarely happen in practice. The dual-polarized system handles the polarization mismatch properly and shows better performance than the mono-polarized system because a dual-polarized array offers the advantage of polarization diversity.

In Figs. 4a and 4b, the data rate of the proposed soft-alignment algorithm is compared with that of the hard-alignment algorithm with multiple round sampling in \cite{Ref_Hur13} against total sounding time $L$. In these simulations, the adaptive channel sounding, which adjusts the sounding time, is not applied to the soft-alignment algorithm. In Figs. 4a and 4b, the simulation results are presented {in different cases of SNR, i.e., $\rho=-8,-10,$ and $-12$ dB.} The data rate of the proposed soft-alignment algorithm increases linearly with $L$ since the effective SNR $|m_{ab}^{s}[q]|^2$ in (\ref{eq:18}), which scales linearly with beam alignment performance, is a function of $L$.  In comparison, the data rate of the hard-alignment algorithm reaches a threshold because of its fundamental limits on $|m_{ab}^{h}[{\ell}]|^2$ in (\ref{eq:13}). At this time, it is interesting to discuss the performance gap between the two beam alignment algorithms when $L=\frac{M_t}{2}$. If $L=\frac{M_t}{2}$, the upper bound of $|m_{ab}^{s}[q]|^2$ for soft-alignment algorithms is the same as that of $|m_{ab}^{h}[{\ell}]|^2$ for hard-alignment algorithms. A large number $Q \geq L$ of candidate codewords that generate narrow beams are used for soft-alignment, while only $L$ codewords are used for hard-alignment. Therefore, $\mathrm{E}\big[|m_{ab}^{s}[q]|^2\big]$ is much bigger than $\mathrm{E}\big[|m_{ab}^{h}[{\ell}]|^2\big]$. Simulation results show that the soft-alignment algorithm scans the channel subspace better than the hard-alignment algorithm.

\begin{table}
\caption{Beam alignment schemes compared in Fig. 4.}
\centering
\begin{tabular}{|c|c|}
  \hline
         & $\rho=[-10,-8,-6,-4,-2, 0, 2]~\mathrm{dB}$~  \\ \hline
Proposed algorithm   & $M_t=32,~L=[ 93~ 59~ 37~ 24~ 16~ 16~ 16]$  \\
 w. adaptive sounding   &    $M_t=64,~L=[ 129~ 81~ 52~ 33~ 32~ 32~ 32]$   \\ \hline
     Hard-alignment  &   $M_t=32,~L=[ 93~ 59~ 37~ 24~ 16~ 16~ 16]$ \\
  w. adaptive sounding &  $M_t=64,~L=[ 129~ 81~ 52~ 33~ 32~ 32~ 32]$  \\ \hline
  Hard-align. w. multiple &   Fixed sounding time  \\
   round sampling \cite{Ref_Hur13}&   $M_t=32,64,~L=60$ \\ \hline
\end{tabular}
\end{table}

\begin{figure}[!t]
\centering
\subfloat[$M_t=32,~ M_r=2$]{\includegraphics[width=0.2441\textwidth]{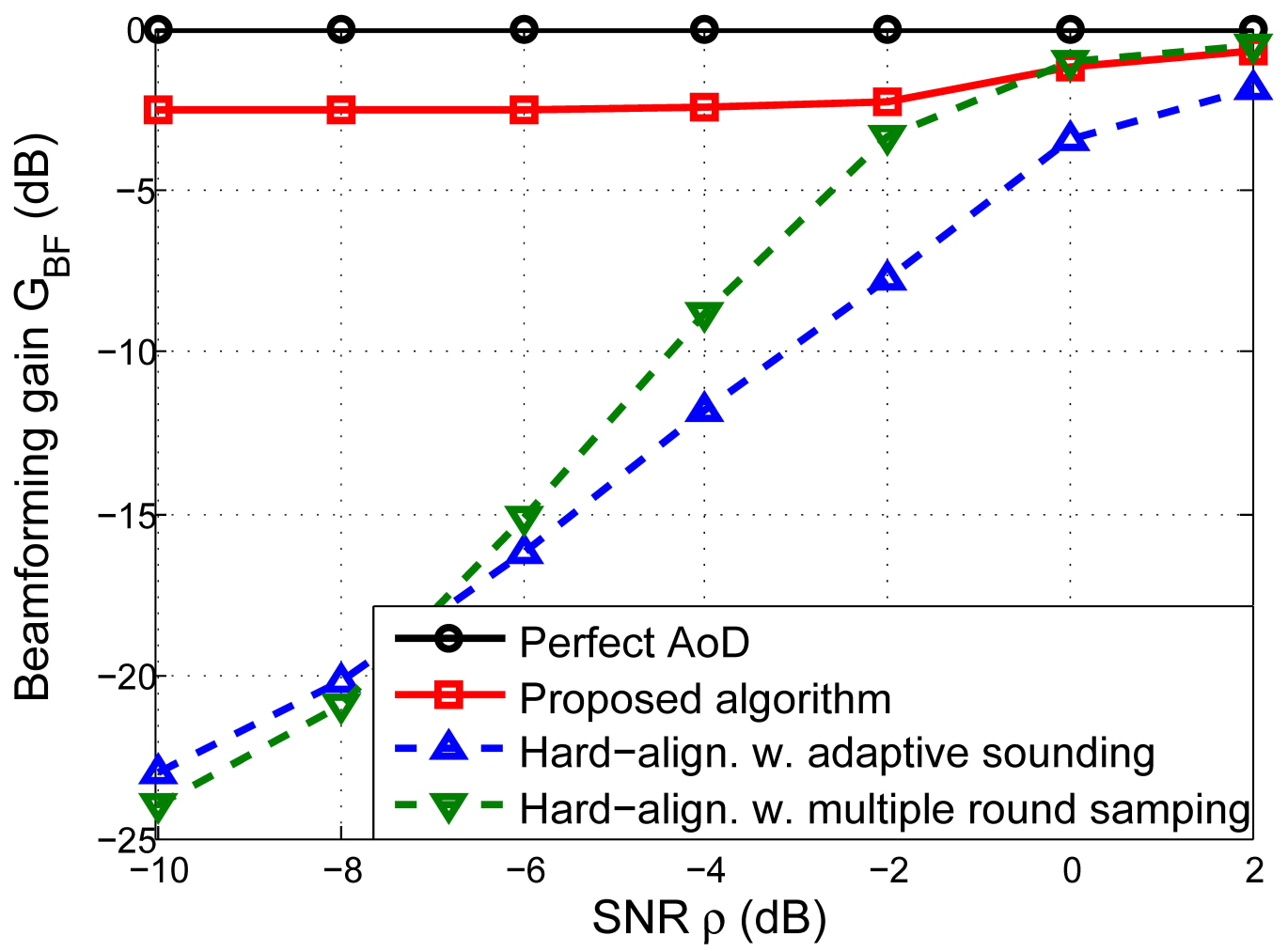}}
\hfil
\subfloat[$M_t=64,~ M_r=2$]{\includegraphics[width=0.2441\textwidth]{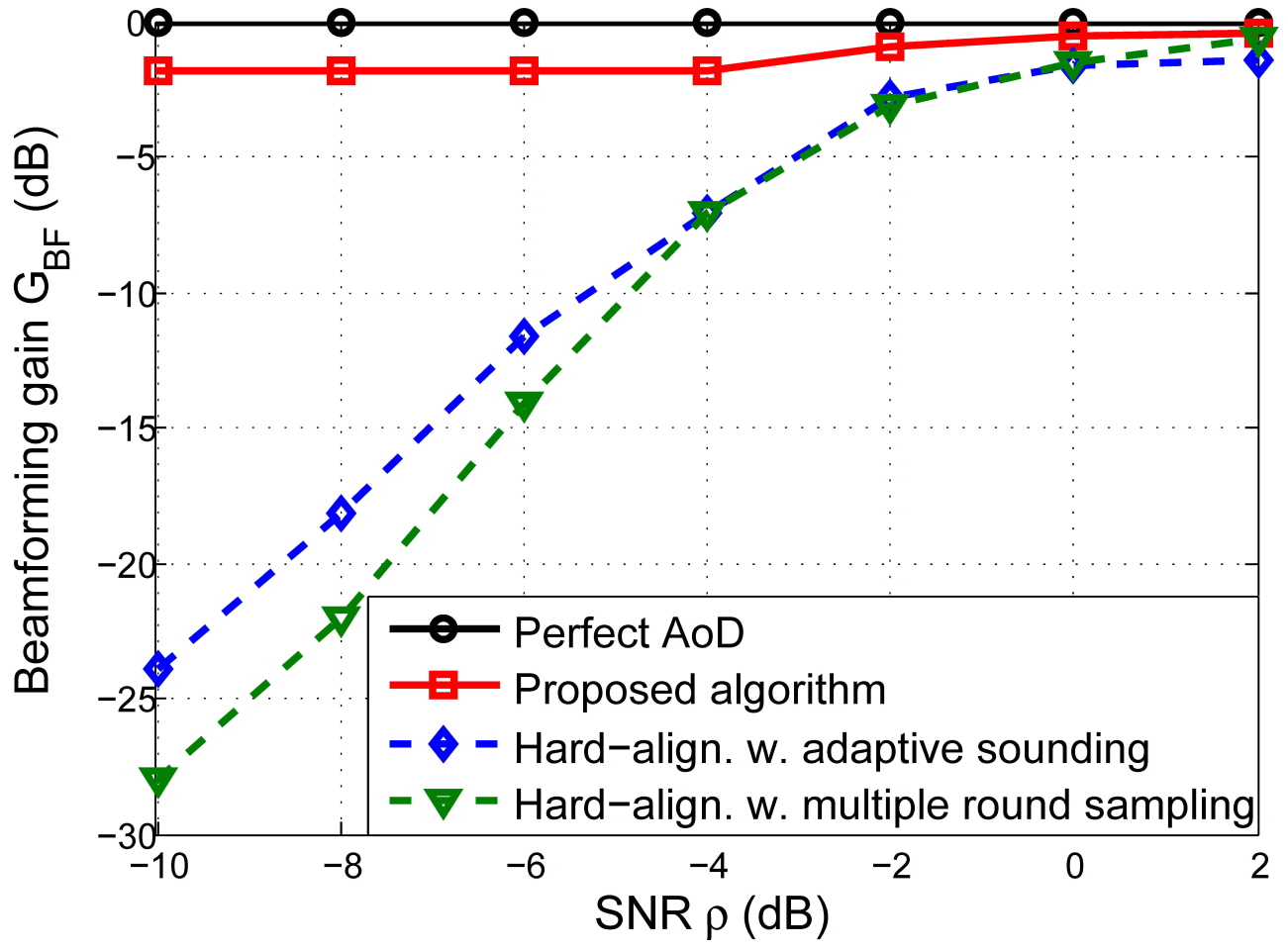}}
\caption{Beamforming gain of proposed and hard beam alignment algorithm against SNR $(M_t=32,~64,$ $M_r=2,$ $\chi=0.2).$}
\end{figure}

In Figs. 5a and 5b, the beamforming gain of the beam alignment algorithms is compared against SNR $\rho$. In the proposed algorithm, the sounding time $L$ is computed based on the inequalities of the adaptive sounding algorithm in (\ref{eq:24}) taking $\rho$, $M_t$, $|\alpha|^2=\frac{1}{2}$, and $\epsilon=0.60$ into account. In Table I, the computed sounding times based on the proposed algorithm and hard-alignment algorithm with adaptive sounding algorithm are summarized as a function of SNR, sorted in ascending order (from $-10~\mathrm{dB}$ to $2~\mathrm{dB}$). It is clear that beamforming gain of the proposed algorithm remains consistent by means of the adaptive sounding algorithm in the low SNR regime. In comparison with the proposed algorithm utilizing same number of sounding times, the hard-alignment algorithm  shows lower performance because of its fundamental limits on $|m_{ab}^{h}[{\ell}]|^2$ in (\ref{eq:13}). In addition, the hard-alignment scheme using fixed sounding time, i.e., $L=60$, also shows lower performance than the proposed algorithms, especially at low SNR.

\begin{figure}[!t]
\centering
\subfloat[$M_t=32,~ M_r=2$]{\includegraphics[width=0.244\textwidth]{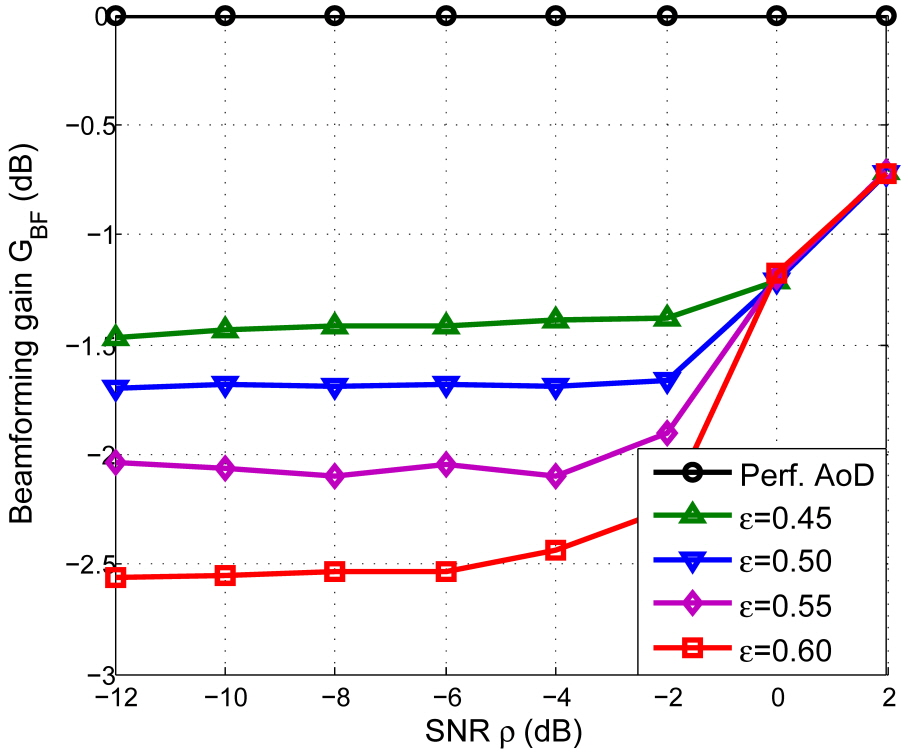}}
\hfil
\subfloat[$M_t=64,~ M_r=2$]{\includegraphics[width=0.244\textwidth]{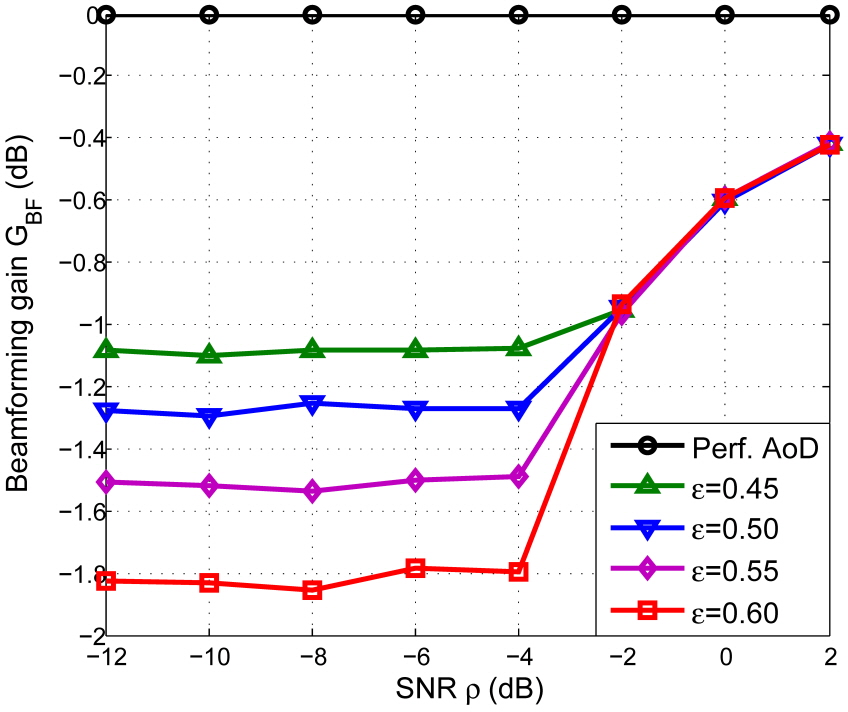}}
\caption{Beamforming gain of proposed algorithm against SNR with different target error probabilities $(M_t=32,~64,$ $M_r=2,$ $\chi=0.2).$}
\end{figure}

In Figs. 6a and 6b, the beamforming gains of the proposed algorithm with various target error probabilities are compared against SNR. In the simulations, target error probabilities $\epsilon$ are set to $0.45$, $0.50$, $0.55$, and $0.60$. Note that sounding time $L$ is computed under different target error probabilities based on the proposed algorithm. It is shown that the beamforming gain performance is maintained according to the predefined target error probabilities. Note that sounding time is restricted by the criterion of the Welch bound $L \geq \frac{M_t}{2}$ for the SNR above $-2~\mathrm{dB}$ in Fig. 6a and $-4~\mathrm{dB}$ in Fig. 6b. Above the SNRs, the final sounding time $L$ is chosen to be $\frac{M_t}{2}$ in different cases of target error probability because $\hat{L}$ in (\ref{eq:24}) is smaller than $\frac{M_t}{2}$. For this reason, all curves are overlapped since each scheme uses same number of sounding times $L=\frac{M_t}{2}$.


\section{Conclusion}
\label{sec:Conclusion}
In this paper, we propose practical beam alignment algorithms for mmWave MIMO systems employing dual-polarized antennas. We first propose a soft-alignment algorithm for a dual-polarized wireless channel. The soft-alignment algorithm scans the channel subspace and relaxes the criterion of the Welch bound with equality training sequences by exploiting polarization diversity. We also propose an adaptive sounding algorithm that selects an efficient amount of sounding time by considering the channel environment. It is shown that the proposed algorithm effectively samples the channel subspace of mobile users even in low SNR environments.


\appendices

\section{Approximated distribution of $\bX$}
\label{sec:A}
We consider the distribution of $\bX$ in (\ref{eq:20}), which is rewritten as
\begin{align*}
\bX= \big(\mathfrak{R}({m}[\check{q}] +  {n}[\check{q}])\big)^2+\big(\mathfrak{I}({m}[\check{q}] +  {n}[\check{q}])\big)^2,
\end{align*}
where $\mathfrak{R}({n}[\check{q}])$ and $\mathfrak{I}({n}[\check{q}])$ follow the normal distribution $\mathcal{N}(0,\frac{1}{2})$. The cdf of $\bX$ is derived as
\begin{align}
\nonumber
\mathrm{F}_{\bX}(x)&=\mathrm{P}(\bX \leq x)
\\
\nonumber
&=\mathrm{P}\big(\big(\mathfrak{R}({m}[\check{q}] +  {n}[\check{q}])\big)^2+\big(\mathfrak{I}({m}[\check{q}] +  {n}[\check{q}])\big)^2 \leq x\big)
\\
\nonumber
&=\int_{-\infty}^{\infty}\mathrm{P}\big((\mathfrak{R}({m}[\check{q}] +  {n}[\check{q}]))^2+(\mathfrak{I}({m}[\check{q}] +  {n}[\check{q}]))^2
\\
\nonumber
&~~~~~~~~~~~~~~\leq x~\big|~ \mathrm{G}[\check{q}]=g\big) {f}_{\bG}(g)dg
\\
\nonumber
&\stackrel{(a)}=\int_{0}^{1} \Big( 1-\textrm{Q}_{1}\big(\sqrt{\rho L |\alpha|^2 g},\sqrt{2x} \big) \Big) {f}_{\bG}(g)dg
\\
\label{eq:exp}
&=\mathrm{E}\big[1-\textrm{Q}_{1}\big(\sqrt{\rho L |\alpha|^2  \bG},\sqrt{2x} \big) \big] \doteq \mathrm{E}\big[U(\bG,x)\big],
\end{align}
where $(a)$ is derived based on the noncentral chi-squared distribution with two degrees of freedom \cite{Ref_Pro01} using $(\mathfrak{R}({m}[\check{q}] ))^2+(\mathfrak{I}({m}[\check{q}] ))^2=|m[\check{q}]|^2=\frac{\rho L}{2}|\alpha|^2\bG$,  defined in (\ref{eq:18}). Note that $\bG \doteq \mathrm{G}[\check{q}]=|\tilde{\bh}^H \bee_{\check{q}} |^2$ is the beamforming gain between the normalized sub-channel and the selected codeword vector.

\begin{figure}[!t]
\centering
\subfloat[$M_t=32$]{\includegraphics[width=0.244\textwidth]{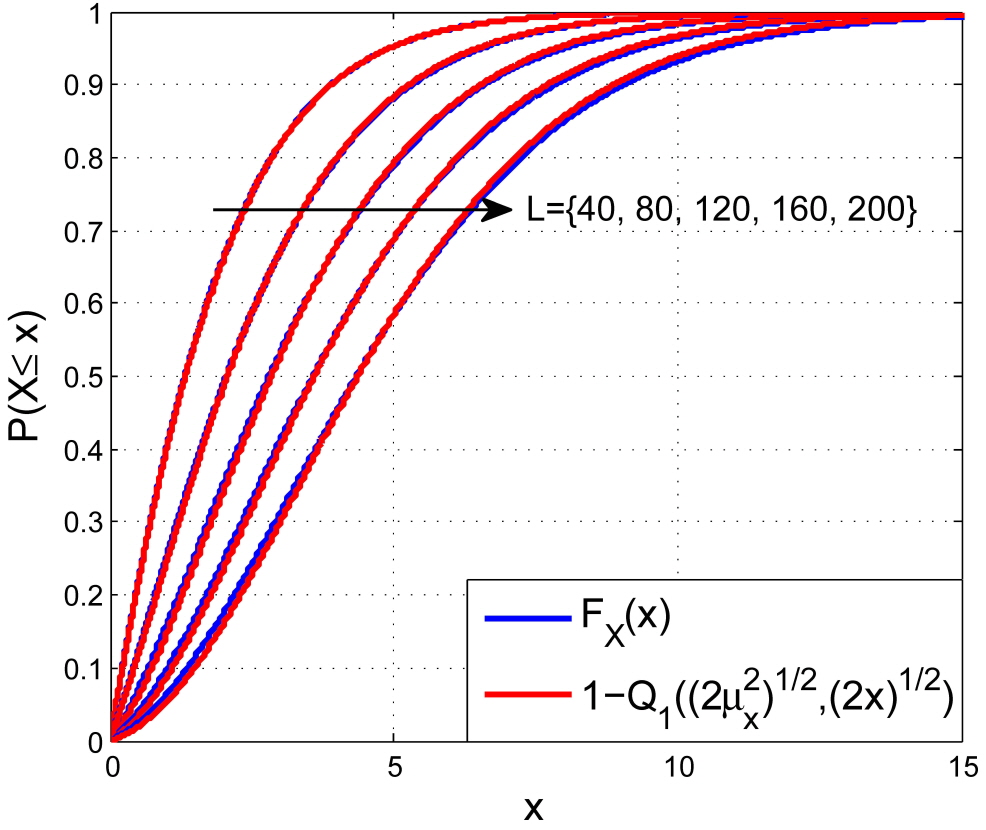}}
\hfil
\subfloat[$M_t=64$]{\includegraphics[width=0.244\textwidth]{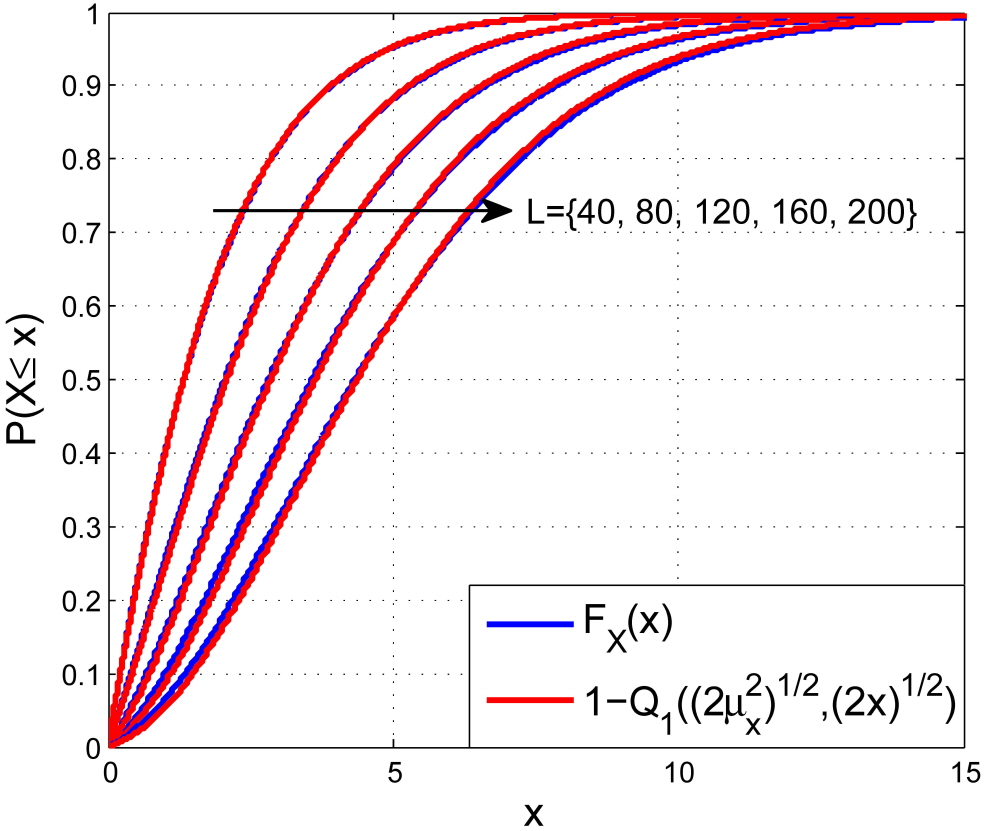}}
\caption{Distribution analysis of $\bX$ $(\mu_x^2=\frac{\rho L}{2}|\alpha|^2 \mathrm{E}[\mathrm{G}[\check{q}] ]$, $\rho=-10$ dB$).$}
\end{figure}

To compute the expectation in (\ref{eq:exp}), the distribution of $\bG$ should be derived first. However, it is difficult to write the distribution of $\bG$ in closed-form. Because it is hard to compute $\mathrm{E}\big[U(\bG,x)\big]$ directly, we derive an approximated cdf of $\bX$. The distribution of $\bG$, considered in Appendix \ref{sec:B} in detail, is negligible outside an interval $\big[\mathrm{E}[\bG]-\delta,\mathrm{E}[\bG]+\delta\big]$ with an arbitrarily small value $\delta$ because it is concentrated near its mean $\mathrm{E}[\bG]$ and bounded by its upper and lower bound. In this case, $\mathrm{E}\big[ U(\bG,x) \big]$ can be approximated based on \cite{Ref_Pap02},
\begin{align*}
\mathrm{F}_{\bX}(x)&=\mathrm{E}\big[ U(\bG,x) \big]
\\
&\simeq U\big(\mathrm{E}[\bG],x \big) \int_{\mathrm{E}[ \bG ]-\delta}^{\mathrm{E}[ \bG ]+\delta}f_{\bG}(g) dg
\\
&\simeq U\big(\mathrm{E}[ \bG],x \big)
\\
&=  \Big( 1-\textrm{Q}_{1}\big(\sqrt{ 2 \mu_x^2},\sqrt{2x} \big) \Big)\mathbbm{1}_{[0,\infty)}(x),
\end{align*}
where $\mu_x^2\doteq \mathrm{E}\big[|m[\check{q}]|^2\big]=\frac{\rho L}{2} |\alpha|^2 \mathrm{E}[\bG]$ denotes the noncentrality parameter of $\bX$.

In Figs. 7a and 7b, the cdf of $\bX$ is compared with the approximated cdf of $\bX$ under $\mu_x^2$ with various sounding times $L$. Note that the approximation holds well when a variance of $\mu_x^2$ is small \cite{Ref_Pap02}. Simulation results show that the approximated cdf approaches the empirical cdf $\mathrm{F}_{\bX}(x)$ in most cases of $\mu_x^2$ in our works.

\section{Expectation of beamforming gain}
\label{sec:B}
In this section, the beamforming gain $\mathrm{G}[{\check{q}}]$ is considered. As mentioned before, we design the codebook $\cE_1$ with $Q_1=\frac{M_t}{2}$ codewords based on \cite{Ref_Hur13,Ref_Hur11}. In the codebook, when $Q_1 = \frac{M_t}{2}$,  each codeword is represented by the normalized array response vector $\sqrt{\frac{2}{M_t}}\ba_{\frac{M_t}{2}}(\theta)$ in (\ref{eq:06}). Thus, the beamforming gain $\mathrm{G}[\check{q}]$ in (\ref{eq:23}) between the normalized sub-channel  $\tilde{\bh}$ and the $\check{q}$-th codeword $\bee_{\check{q}}=\frac{1}{\sqrt{N}}\ba_{N}(\theta_{\check{q}})$ with $N$ elements is given by \cite{Ref_Han98}
\begin{align}
\label{eq:25}
\mathrm{G}[\check{q}]&=\frac{|\ba_{N}^{H}(\theta_{\check{q}}+\phi_{\check{q}})  \ba_{N}(\theta_{\check{q}})|^2 }{N^2}=\frac{1}{N^{2}} \frac{\sin^{2}(\frac{\pi \eta N}{2})}{\sin^{2}(\frac{\pi \eta }{2})} \doteq \Gamma(|\eta|)
\end{align}
where $\theta_{\check{q}}$ is the AoD beam direction of the $\check{q}$-th codeword, $\phi_{\check{q}}$ is the beam direction difference between the $\check{q}$-th codeword and the sub-channel vector and $\eta \doteq \sin\theta_{\check{q}}-\sin(\theta_{\check{q}}+\phi_{\check{q}})$ is the beam direction difference in the $\psi$-domain in Fig. 8. $\mathrm{G}[\check{q}]$  is bounded by its upper and lower bound, i.e., $1$ and $\tau^2$.

\begin{figure}
\centering
\includegraphics[width=0.490\textwidth]{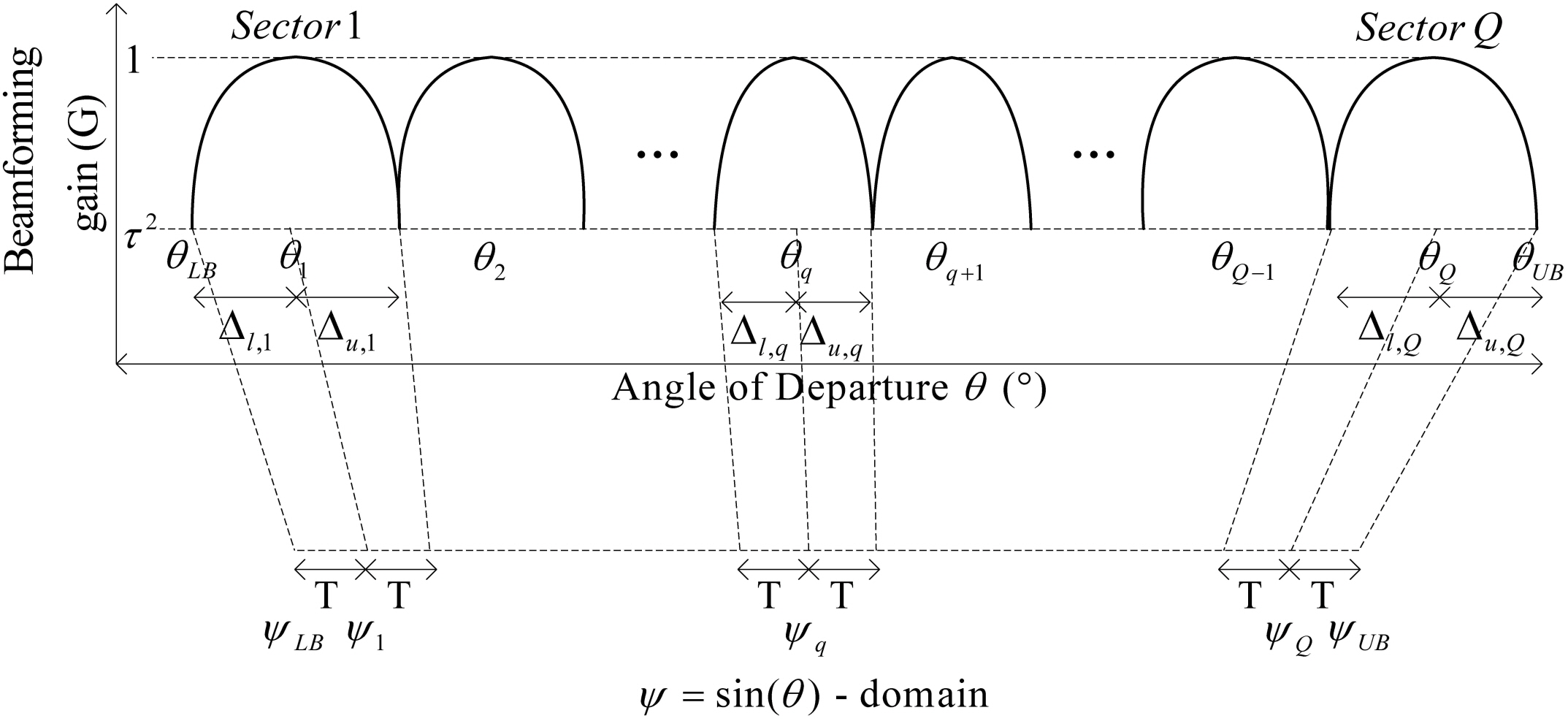}
\caption{Quantized sector for each codeword.}
\end{figure}

In the codebook of \cite{Ref_Hur13,Ref_Hur11}, the beam-width of each codeword is considered in the $\psi$-domain. To optimize the codebook, the beam-widths of each codeword is equally divided in the $\psi$-domain as $\mathrm{2T} \doteq \frac{\sin\theta_{UB}-\sin\theta_{LB}}{Q}$  where $\theta_{UB}$ and $\theta_{LB}$ are the upper and lower bounds of the entire range, respectively. Then, the upper beam-width $\Delta_{u,\check{q}}$ and the lower beam-width $\Delta_{l,\check{q}}$ of each codeword is given by
\begin{align}
\label{eq:26}
\sin(\theta_{\check{q}}+\Delta_{u,{\check{q}}})-\sin\theta_{\check{q}}&= \mathrm{T},
\\
\label{eq:27}
\sin\theta_{\check{q}}-\sin(\theta_{\check{q}}+\Delta_{l,{\check{q}}})&= \mathrm{T}.
\end{align}

In the ${\check{q}}$-th quantized sector, $\Delta_{u,{\check{q}}}$ and $\Delta_{l,{\check{q}}}$  are dependent on the beam direction $\theta_{\check{q}}$ of the codeword $\bee_{\check{q}}$. In order to compute the expected value of the beamforming gain, the beam-width of each sector needs to be represented by the beam direction $\theta_q$. The beam-width of each codeword in the $\theta$-domain is approximated in the following lemma.

\newtheorem{lemma}{Lemma}
\begin{lemma}The upper and lower beam-widths of each codeword may be approximated as
\end{lemma}
\begin{align}
\nonumber
\Delta_{\theta_q} \doteq \Delta_{u,q}=\Delta_{l,q} \simeq \frac{ \mathrm{T}}{\cos\theta_q},
\end{align}
when the beam-width is sufficiently small.
\begin{IEEEproof}
First, we consider the upper beam-width of $q$-th sector with $\bee_q$. The upper beam-width $\Delta_{u,q}$  is derived as
\begin{align}
\nonumber
\mathrm{T}&=\sin\theta_q\cos(\Delta_{u,q})+\cos\theta_q\sin(\Delta_{u,q})-\sin\theta_q
\\
\nonumber
&\simeq \sin\theta_q\Big(1-\frac{\Delta_{u,q}^2}{2}\Big)+\cos\theta_q\Delta_{u,q}-\sin\theta_q
\\
\nonumber
&\simeq \cos\theta_q\Delta_{u,q}.
\end{align}
The first approximation is based on the small-angle approximation of trigonometric functions, i.e., $\sin\Delta \simeq \Delta,~ \cos\Delta \simeq 1-\frac{\Delta^2}{2}$, since the half beam-width $\Delta_{u,q}$ is designed to have a sufficiently small value. Then, we drop the $\frac{\Delta_{u,q}^2}{2}$ term since $\Delta_{u,q}^2$ approaches $0$.

The lower beam-width in (\ref{eq:27}) is approximated in the same way and has an identical value. The upper and lower beam-width of the $q$-th sector in the $\theta$-domain is approximated as
\begin{align*}
&\Delta_{\theta_q} \doteq \Delta_{u,q} = \Delta_{l,q} \simeq \frac{\mathrm{T}}{\cos\theta_q}.
\end{align*}
The upper and lower beam-width are functions of the beam direction $\theta_q$ and the half beam-width $\mathrm{T}$ in the $\psi$-domain.
\end{IEEEproof}

With the approximate beam-width in Lemma $1$, the expectation of beamforming gain $\mathrm{E}[\Gamma(|\eta|)]$ is upper bounded. The beamforming gain is defined with two random variables, $\theta_{\check{q}}$ and $\phi_{\check{q}}$ in (\ref{eq:25}). Because it is hard to derive $\mathrm{E}[\Gamma(|\eta|)]$ directly, we derive an upper bound of $\mathrm{E}[\Gamma (|\eta|)]$.
\begin{lemma} An upper bound of $\mathrm{E}[\Gamma(|\eta|)]$ is
\end{lemma}
\begin{align*}
\mathrm{E}\big[\mathrm{G}[{\check{q}}]\big]=\mathrm{E}\big[\Gamma(|\eta|)\big] \leq \Gamma\big(\mathrm{E}[|\eta|]\big) \simeq \Gamma\Big(\frac{\mathrm{T}}{2}\Big)  \doteq \varsigma^2.
\end{align*}
\begin{IEEEproof}
 In this proof, we drop the codeword index term ${\check{q}}$ for simplicity. The upper bound of $\mathrm{E}[\Gamma(|\eta|)]$ is computed based on Jensen's inequality since $\Gamma(|\eta|)$ is a concave function with regard to $|\eta|$. $|\eta|$, which represents the beam direction difference between two array manifold vectors, is written as
\begin{align*}
&|\eta| =\left\{\begin{array}{ll}  \sin(\theta+\phi)-\sin\theta, & 0 \leq \phi \leq \Delta_{\theta}  \\ \sin\theta-\sin(\theta+\phi) , & -\Delta_{\theta} \leq \phi < 0 \end{array}\right.
\end{align*}
where $\Delta_{\theta} \simeq \frac{\mathrm{T}}{\cos\theta}$ is the approximate half beam-width which is defined in Lemma $1$. We assume $\phi \sim \mathrm{U}\big(-\Delta_{\theta}~\Delta_{\theta}\big)$.

First, we compute the expected value of $\sin(\theta+\phi)-\sin\theta$ when $0 \leq \phi \leq \Delta_{\theta}$. The expected value is derived as follows
\begin{align}
\nonumber
&\mathrm{E}_{\theta,\phi}[\sin(\theta+\phi)-\sin\theta]
\\
\nonumber
&=\int_{-\theta_{UB}}^{\theta_{UB}} \bigg({\int_{0}^{\Delta_{\theta}}     { \big(\sin(\theta+\phi)-\sin\theta\big) f(\phi) d\phi} \bigg)         f(\theta) d \theta     }
\\
\nonumber
&=\int_{-\theta_{UB}}^{\theta_{UB}}     {   \bigg(  \frac{2}{\Delta_{\theta}}\sin(\theta+\frac{\Delta_{\theta}}{2})\sin(\frac{\Delta_{\theta}}{2})-\sin\theta \bigg)     f(\theta) d \theta }
\\
\nonumber
&\simeq \int_{-\theta_{UB}}^{\theta_{UB}}        {   \bigg( \sin(\theta+\frac{\Delta_{\theta}}{2}) -\sin\theta \bigg)     f(\theta) d \theta }
\\
\nonumber
&=\frac{1}{{2\theta_{UB}} } \int_{-\theta_{UB}}^{\theta_{UB}}        { \bigg( \sin\theta\big(\cos(\frac{\Delta_{\theta}}{2})-1\big)+\cos\theta\sin(\frac{\Delta_{\theta}}{2})    \bigg)  d \theta }
\\
\nonumber
&\simeq \frac{1}{{2\theta_{UB}} }\int_{-\theta_{UB}}^{\theta_{UB}}        { \bigg( { \sin\theta\frac{(\frac{\Delta_{\theta}}{2})^2}{2}+\cos\theta \frac{\Delta_{\theta}}{2} }   \bigg)  d \theta }
\\
\nonumber
&\simeq \frac{1}{{2\theta_{UB}} }\int_{-\theta_{UB}}^{\theta_{UB}}        {  { \cos\theta \frac{\mathrm{T}}{2\cos\theta} }     d \theta }=\frac{\mathrm{T}}{2}.
\end{align}

The first and second approximations are based on the small angle approximation technique for trigonometric functions, i.e., $\sin(\frac{\Delta_{\theta}}{2}) \simeq \frac{\Delta_{\theta}}{2},~\cos(\frac{\Delta_{\theta}}{2}) \simeq 1-\frac{(\frac{\Delta_{\theta}}{2})^2}{2}$,   when $\frac{\Delta_{\theta}}{2}$ is sufficiently small. In the third approximation, the term $\frac{(\frac{\Delta_{\theta}}{2})^2}{2}$ is dropped since $(\frac{\Delta_{\theta}}{2})^2$ approaches $0$. In the last approximation, $\Delta_{\theta}$ is replaced by $\frac{\mathrm{T}}{\cos\theta}$ which is defined in Lemma 2.

The approximation of $\mathrm{E}[\sin\theta-\sin(\theta-\phi)]$ when $-\Delta_{\theta} \leq \phi < 0$ is derived in the same way and it is identical with that of $\mathrm{E}[\sin(\theta+\phi)-\sin\theta]$. The expectation of $|\eta|$ is then
\begin{align*}
\mathrm{E}[|\eta|] \simeq \frac{\mathrm{T}}{2}.
\end{align*}
\end{IEEEproof}

\begin{figure}[!t]
\centering
\subfloat[$N=16,$ $\theta_{UB}=-\theta_{LB}=\frac{\pi}{3}$]{\includegraphics[width=0.244\textwidth]{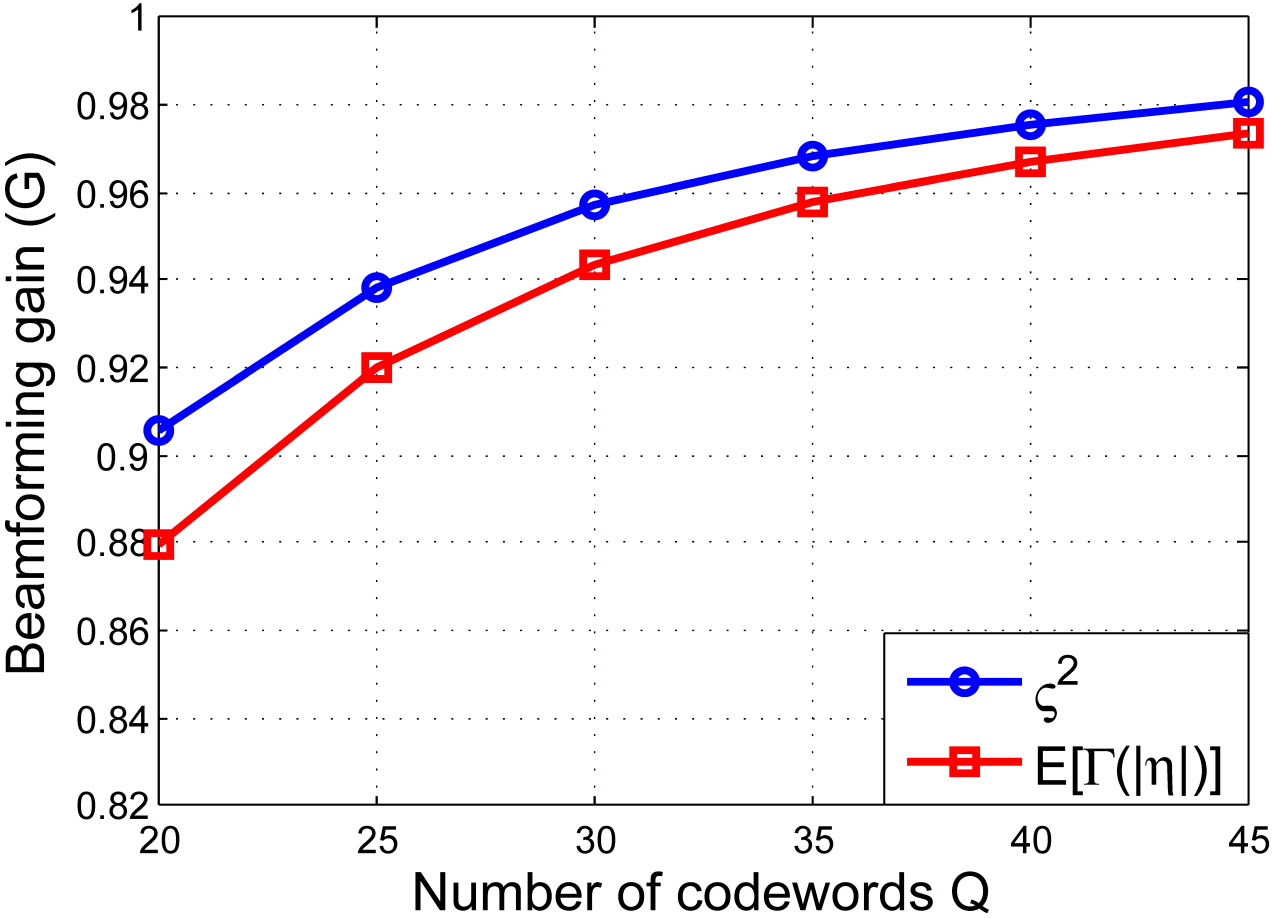}}
\hfil
\subfloat[$N=32,$ $\theta_{UB}=-\theta_{LB}=\frac{\pi}{3}$]{\includegraphics[width=0.244\textwidth]{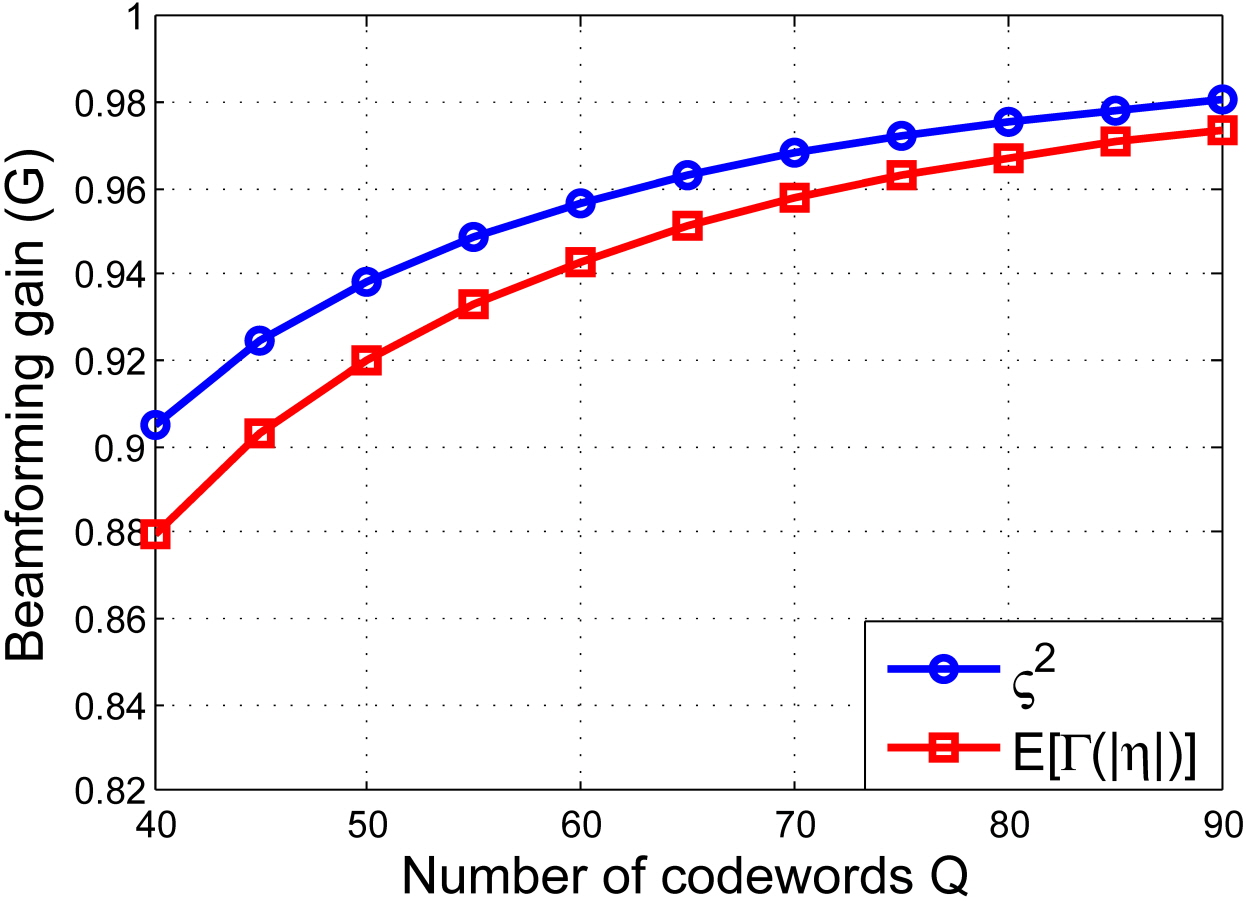}}
\hfil
\subfloat[$N=16,$ $Q=32,$ $\theta_{UB}=-\theta_{LB}$]{\includegraphics[width=0.244\textwidth]{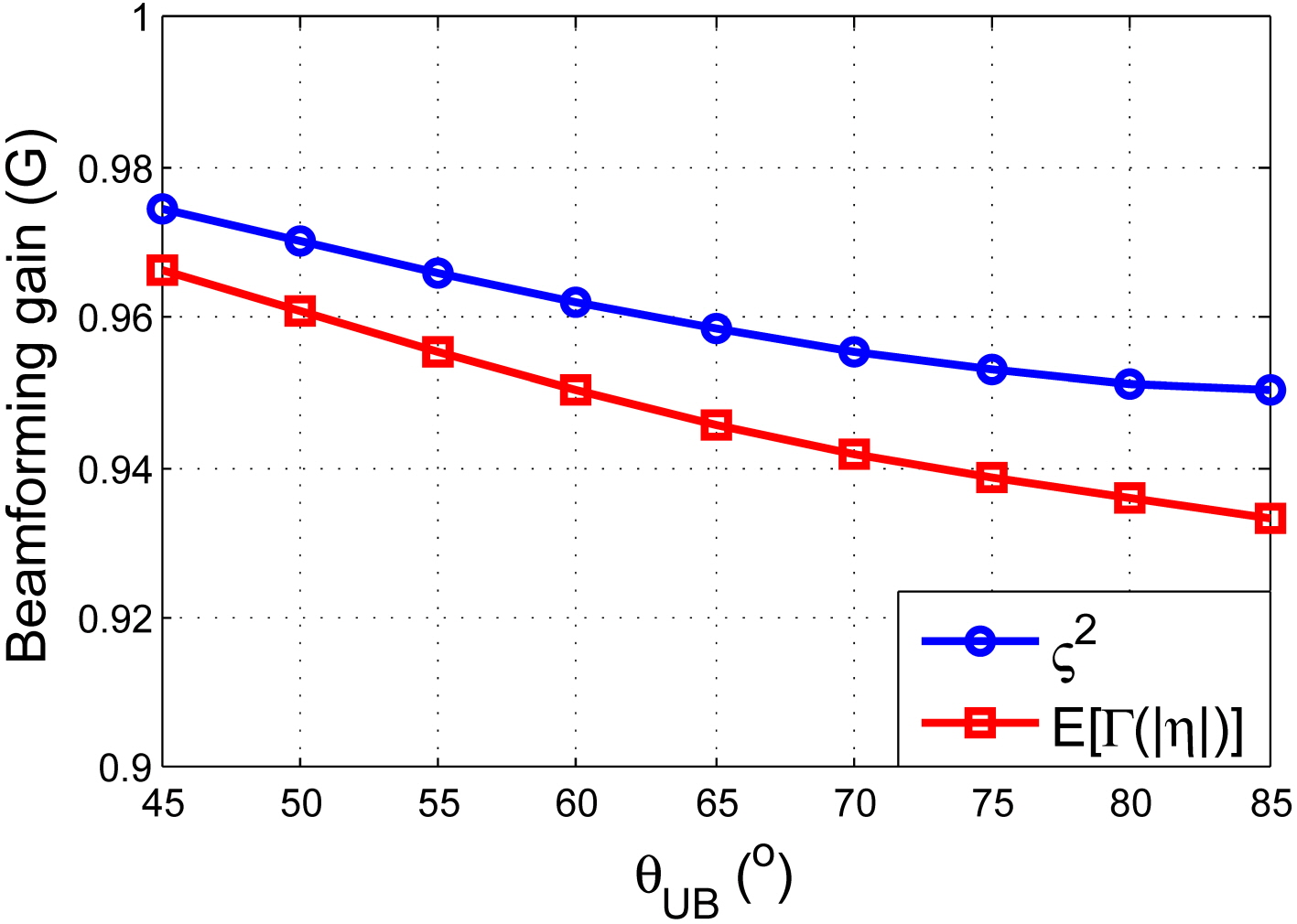}}
\hfil
\subfloat[$N=32,$ $Q=64,$ $\theta_{UB}=-\theta_{LB}$]{\includegraphics[width=0.244\textwidth]{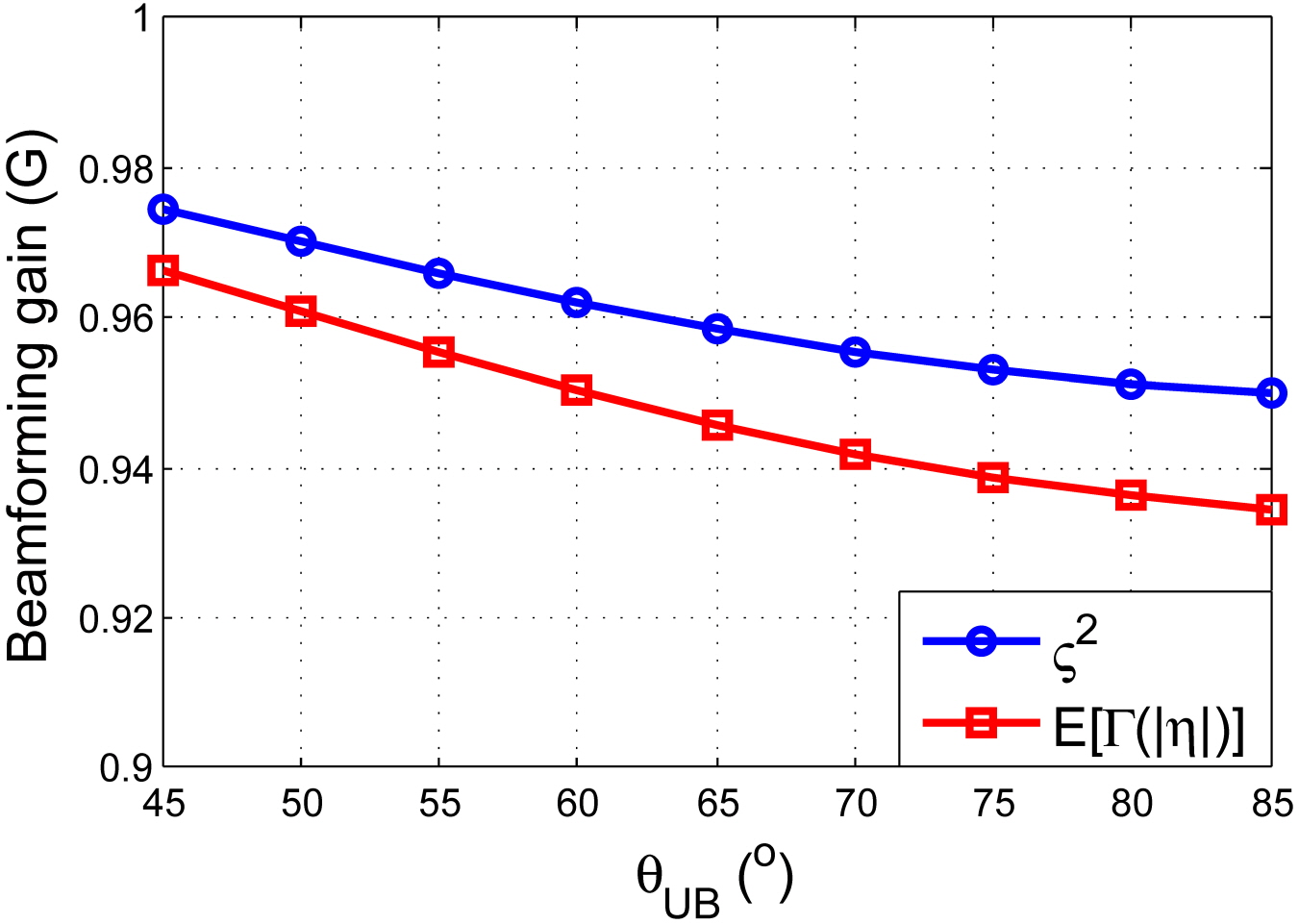}}
\caption{Beamforming gain analysis of codebook \cite{Ref_Hur13, Ref_Hur11}.}
\end{figure}

In Fig. 9, the approximate upper bound of $\mathrm{E}[\Gamma(|\eta|)]$ is compared with the simulation results with the codebook. In Figs. 9a and 9b, the beamforming gain of the codebook is compared against the number of codewords under $\theta_{UB}=-\theta_{LB}=\frac{\pi}{3}$. In Figs. 9c and 9d, the beamforming gain of the codebook is compared against the bounds of possible AoDs. Note that the simulation results for $\mathrm{E}[\Gamma(|\eta|)]$ are upper bounded by the formulation $\varsigma^2$ in Lemma $2$.



\bibliographystyle{IEEEtran}
\bibliography{refs_BA}







\end{document}